\documentclass[twocolumn,pre,aps,showpacs,showkeys,amsmath]{revtex4}
\usepackage{graphicx}
\usepackage{adjustbox}
\usepackage{multirow}

\begin{document}

\title{Dynamical Origin for Winner-Take-All Competition in A Biological Network of The Hippocampal Dentate Gyrus}
\author{Sang-Yoon Kim}
\email{sykim@icn.re.kr}
\author{Woochang Lim}
\email{wclim@icn.re.kr}
\affiliation{Institute for Computational Neuroscience and Department of Science Education, Daegu National University of Education, Daegu 42411, Korea}

\begin{abstract}
We consider a biological network of the hippocampal dentate gyrus (DG). The DG is a pre-processor for pattern separation which facilitates pattern storage and retrieval in the CA3 area of the hippocampus. The main encoding cells in the DG are the granule cells (GCs) which receive the input from the entorhinal cortex (EC) and send their output to the CA3. We note that the activation degree of GCs is so low ($\sim 5 \%$). This sparsity has been thought to enhance the pattern separation. We investigate the dynamical origin for winner-take-all (WTA) competition which leads to sparse activation of the GCs.
The whole GCs are grouped into lamellar clusters. In each GC cluster, there is one inhibitory (I) basket cell (BC) along with excitatory (E) GCs.
There are three kinds of external inputs into the GCs; the direct excitatory EC input, the indirect inhibitory EC input, mediated by the HIPP
(hilar perforant path-associated) cells, and the excitatory input from the hilar mossy cells (MCs). The firing activities of the GCs are determined via competition between the external E and I inputs. The E-I conductance ratio ${ {\cal{R}}_{\rm E-I}^{\rm (con)}}^*$ (given by the time average of the external E to I conductances) may represents well the degree of such external E-I input competition. It is thus found that GCs become active when their
${ {\cal{R}}_{\rm E-I}^{\rm (con)}}^*$ is larger than a threshold ${\cal{R}}_{th}^*$, and then the mean firing rates of the active GCs are strongly correlated with ${ {\cal{R}}_{\rm E-I}^{\rm (con)}}^*$. In each GC cluster, the feedback inhibition of the BC may select the winner GCs. GCs with larger
${ {\cal{R}}_{\rm E-I}^{\rm (con)}}^*$ than the threshold ${\cal{R}}_{th}^*$ survive, and they become winners; all the other GCs with smaller
${ {\cal{R}}_{\rm E-I}^{\rm (con)}}^*$ become silent. In this way, WTA competition occurs via competition between the firing activity of the GCs and the feedback inhibition from the BC in each GC cluster. In this case, the hilar MCs are also found to play an essential role of enhancing the WTA competition in each GC cluster by exciting both the GCs and the BC.
\end{abstract}


\maketitle

\section{Introduction}
\label{sec:INT}
The hippocampus, consisting of the dentate gyrus (DG) and the areas CA3 and CA1, is known to play a key role in memory formation, storage, and retrieval
(e.g., episodic memory involving an arbitrary association between events characterizing an episode) \cite{Gluck,Squire}. In this hippocampus, the area CA3
has been often considered to operate as an autoassociation network, because there are extensive recurrent collateral synapses between the pyramidal cells
in the CA3. The autoassociation network stores input ``patterns'' in modifiable synapses between the pyramidal cells. Then, when a partial or noisy version
of the stored pattern is presented, activity of pyramidal cells propagates along the previously-strengthened pathways and reinstates the complete stored pattern,
which is called the process of pattern completion. The idea of hippocampal autoassociation network has originated in the work of Marr \cite{Marr,Will} and
was elaborated later by many others \cite{Mc,Rolls1,Rolls2a,Rolls2b,Treves1,Treves2,Treves3}.

Storage capacity of autoassociation memory corresponds to the number of distinct patterns that can be stored and recalled.
It may be increased if the input patterns are sparse (i.e., there are only a few active elements in each pattern) and non-overlapping/orthogonalized
[i.e., active elements in one pattern may be likely to be (inactive) silent elements in other patterns]
\cite{Marr,Will,Mc,Rolls1,Rolls2a,Rolls2b,Treves1,Treves2,Treves3,Oreilly}. This process of transforming a set of input patterns
into sparser and orthogonalized patterns is called pattern separation.

The DG is the first subregion of the hippocampus that receives inputs from the entorhinal cortex (EC) via the perforant paths (PPs).
As a pre-processor for the CA3, the primary granule cells (GCs) in the DG performs pattern separation on the input patterns coming from the EC
by sparsifying and orthogonalizing them (i.e., the input patterns from the EC become sparser and orthogonalized via pattern separation of the GCs
in the DG) \cite{Treves3,Oreilly,Schmidt,Rolls3,Knier,Myers1,Myers2,Scharfman,Chavlis,Yim,PS1,PS2,PS3,PS4,PS5,PS6,PS7}. Then, the pattern-separated outputs are projected (from the GCs) to the pyramidal cells in the CA3 via the mossy fibers (MFs). These sparse, but relatively strong MFs play a role of ``teaching inputs'' (to the autoassociation network in the CA3) which tend to trigger synaptic plasticity between the pyramidal cells and also between the pyramidal cells and the EC cells \cite{Treves3,Oreilly,Rolls3,Myers2,Scharfman}. Thus, a new pattern may be stored in modified synapses (i.e., pattern storage may occur via synaptic plasticity caused by the MFs). In this way, pattern separation in the DG facilitates pattern storage in the CA3.

In addition to the indirect inputs from the EC to the CA3 through the DG (i.e., the projections of the outputs from the DG onto the CA3 via the MFs are responsible for pattern storage), direct weaker inputs from the EC to the CA3 pyramidal cells via PPs represent partial or noisy version of patterns to be recalled.
These direct EC inputs  would activate a subset of pyramidal cells in the CA3 which would in turn activate other pyramidal cells through the previously-strengthened
synapses until the complete stored pattern is recalled (i.e., stored patterns may be recalled via pattern completion)
\cite{Treves3,Oreilly,Rolls3,Myers2,Scharfman}. In this way, the direct EC inputs to the CA3 play a role of retrieval cue for recalling the previously-stored patterns via pattern completion, in contrast to the indirect EC inputs via the MFs from the DG which cause synaptic plasticity leading to pattern storage \cite{Lee}.

In this paper, we consider a biological network of the hippocampal DG. The primary GCs in the DG network receive the input patterns from the EC via the PPs,
perform pattern separation on the EC inputs, and project their outputs onto the CA3 via MFs. In this process of pattern separation, the activation degree $D_a$ of the GCs is so low ($D_a \sim 5$ $\%$). The GCs exhibit sparse firing activity via competitive learning \cite{Rolls1,Treves3,Rolls3,Myers1}, and
the sparsity has been considered to enhance the pattern separation \cite{Treves3,Oreilly,Schmidt,Rolls3,Myers1,Myers2,Scharfman,Chavlis,Yim}.

Here, we investigate the dynamical origin of the winner-take-all (WTA) competition which leads to sparse activation of the GCs (improving the pattern separation)
\cite{WTA1,WTA2,WTA3,WTA4,WTA5,WTA6,WTA7,WTA8,WTA9,WTA10}. We first note that the whole GCs are grouped into the lamellar clusters \cite{Cluster1,Cluster2,Cluster3,Cluster4}. In our DG network, there are 100 (non-overlapping) GC clusters. Each GC cluster consists of 20 excitatory (E) GCs along
with one inhibitory (I) basket cell (BC). Thus, the GCs and the BC in the GC cluster form an E-I dynamical loop where all the GCs are coupled to the single BC; there are no couplings between the GCs. Hence, all the GCs provide excitation to the BC which then gives back the feedback inhibition to all the GCs.
Then, competition between the firing activity of the GCs and the feedback inhibition of the BC selects which GCs fire. Strongly active GCs survive under the feedback inhibition of the BC (i.e., they become winners), while weakly active GCs become silent in response to the feedback inhibition of the BC.

The firing activities of the GCs are determined via competition between the external E and I inputs to the GCs.
The EC is the main external input source for the GCs. There are the direct excitatory EC input via the PPs and the indirect disynaptic inhibitory EC input, mediated
by the HIPP (hilar perforant path-associated) cells in the hilus of the DG (i.e., EC $\rightarrow$ HIPP cell $\rightarrow$ GC).

In the DG, the hilus, consisting of the inhibitory HIPP cells and the excitatory mossy cells (MCs), underlies the GC layer (composed of GCs and BCs)
\cite{Myers1,Chavlis,Yim,Hilus1,Hilus2,Hilus3,Hilus4,Hilus5,Hilus6,Hilus7}. Thus, there are two excitatory cells (i.e., GCs and MCs) in the DG rather than one
in the CA3 and the CA1. The MCs enhances the firing activity of the GC-BC loop in the GC cluster by providing excitation to both the GCs and the BC. Thus, there appears a 3rd type of excitatory input from the hilar MCs into the GCs, in addition to the two kinds of external EC inputs (i.e., the direct excitatory EC input and the indirect inhibitory EC input, mediated by the HIPP cells). Consequently, there are three kinds of external inputs into the GCs; two types of excitatory inputs from the EC via PPs and from the MCs and one kind of inhibitory input from the HIPP cells.

For characterization of the degree of the external E-I input competition, we introduce the E-I conductance ratio ${ {\cal{R}}_{\rm E-I}^{\rm (con)}}^*$,
given by the time average of the ratio of the external E to I conductances, $\overline {g_{\rm E}(t)/g_{\rm I}(t)}$ (the overline denotes time average); the excitatory conductance $g_{\rm E}(t)= g_{\rm EC}(t) + g_{\rm MC}(t)$ ($g_{\rm EC}(t)$: conductance of the excitatory EC input and $g_{\rm MC}(t)$: conductance of the excitatory MC input) and the inhibitory conductance $g_{\rm I}(t)= g_{\rm HIPP}(t)$ (conductance of the inhibitory HIPP input).
When their ${ {\cal{R}}_{\rm E-I}^{\rm (con)}}^*$ is greater than a threshold ${\cal R}_{th}^*$, GCs become active; otherwise, they become silent.
The mean firing rates (MFRs) of the active GCs are also found to be strongly correlated with ${ {\cal{R}}_{\rm E-I}^{\rm (con)}}^*$ (i.e., with increasing
${ {\cal{R}}_{\rm E-I}^{\rm (con)}}^*$, their MFRs also increase). In this way, the degree of the firing activity of the GCs may be well characterized in terms of their ${ {\cal{R}}_{\rm E-I}^{\rm (con)}}^*$.

Then, the feedback inhibition from the BC selects the winner GCs in each GC cluster. GCs with larger ${ {\cal{R}}_{\rm E-I}^{\rm (con)}}^*$ than the threshold
${\cal R}_{th}^*$ are found to survive under the feedback inhibition, and they become winners, while all the other GCs with smaller ${ {\cal{R}}_{\rm E-I}^{\rm (con)}}^*$ become silent in response to the feedback inhibition. Thus, WTA competition occurs through competition between the firing activity of the GCs and the feedback inhibition of the BC.

Finally, we study the WTA competition by changing the fraction of MCs, $F_{\rm MC}$ (i.e., ablating a subset of MCs). With decreasing $F_{\rm MC}$, WTA competition is found to become weaker, and hence more winners appear. In this way, the MCs play an important role to enhance the WTA competition in the GC cluster by exciting the GCs and the BC in the GC-EC loop.

This paper is organized as follows. In Sec.~\ref{sec:DGN}, we describe a biological network of the hippocampal dentate gyrus.
Then, in the main Sec.~\ref{sec:DO}, we investigate dynamical origin for the WTA competition. Finally, we give summary and discussion in Sec.~\ref{sec:SUM}.

\begin{figure}[t]
\includegraphics[width=\columnwidth]{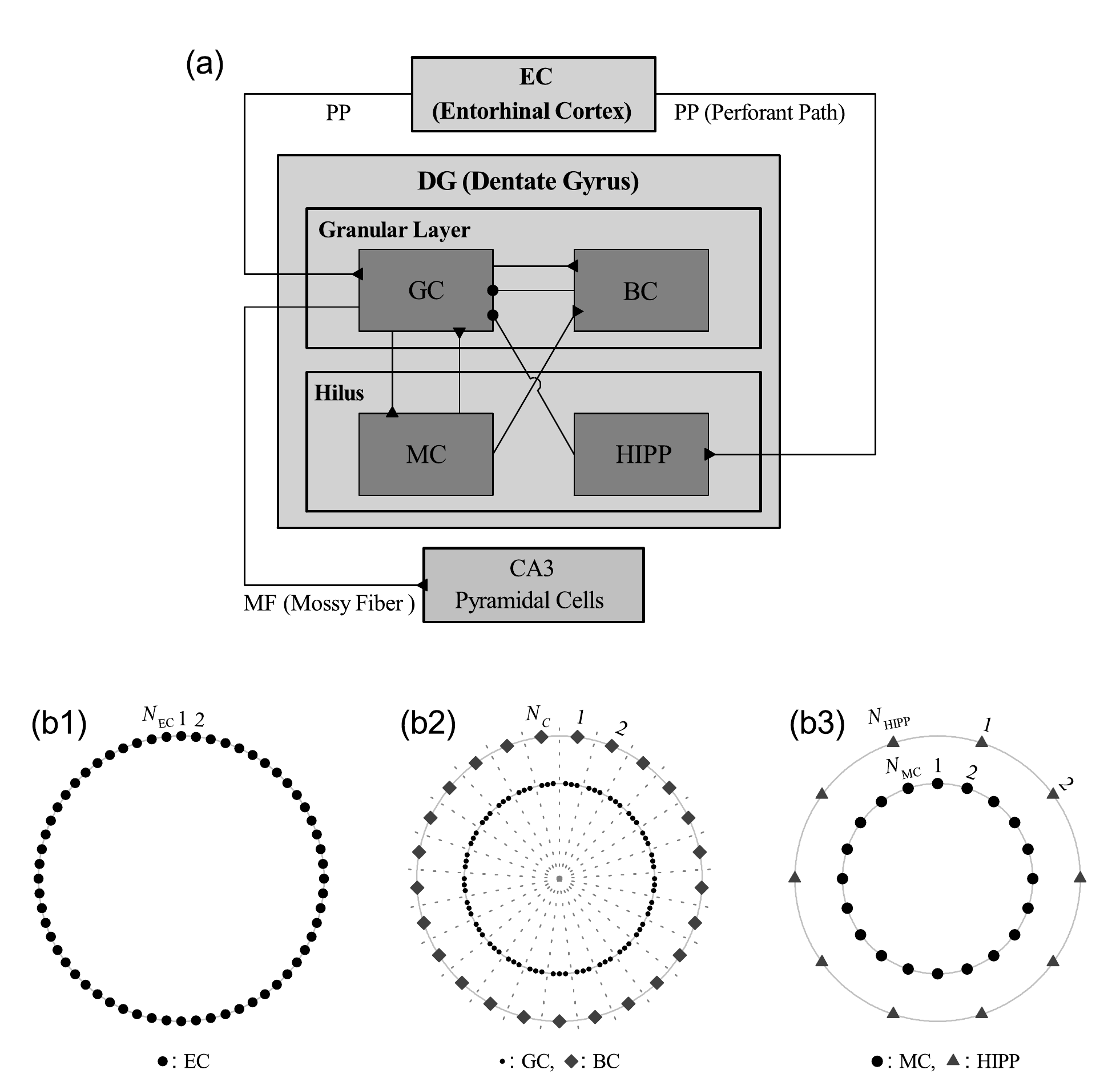}
\caption{Hippocampal dentate gyrus (DG) network. (a) Box diagram for the hippocampal DG network. Lines with triangles and circles denote excitatory and inhibitory synapses, respectively. GC (granule cell) and BC (basket cell) in the granular layer, HIPP cell (hilar perforant path-associated cell) and MC (mossy cell) in the hilus, and EC (entorhinal cortex). Three kinds of ring networks in (b1)-(b3). (b1) Schematic diagram for the EC ring network, composed of $N_{\rm EC}$ EC cells (solid circles). (b2) Schematic diagram for the granular-layer ring network with concentric inner GC and outer BC rings. Numbers represent GC clusters (bounded by dotted lines). Each GC cluster ($I=1,\dots,N_c$) consists of $n_{\rm {GC}}^{(c)}$ GCs (solid circles) and one BC (diamonds). (b3) Schematic diagram for the hilar ring network with concentric inner MC and outer HIPP rings, consisting of $N_{\rm MC}$ MCs and  $N_{\rm HIPP}$ HIPP cells, respectively.
}
\label{fig:DGN}
\end{figure}

\section{Biological Network of The Hippocampal Dentate Gyrus}
\label{sec:DGN}
In this section, we describe our biological network of the hippocampal DG, and present the governing equations for the population dynamics.
Hyperexcitability of the GCs via sprouting of the MFs has been studied in large-scale biological networks of the DG with a high degree of anatomical and physiological realism \cite{BN1,BN2}. These biological networks consist of the hilar cells (e.g., MCs and HIPP cells) as well as the GCs, in contrast to the prior abstract computational models which focused primarily on the GCs (without considering the hilar cells) and performed pattern separation and completion \cite{Marr,Will,Mc,Rolls1,Rolls2a,Rolls2b,Treves1,Treves2,Treves3,Oreilly}. On the other hand, the above works in the biological networks did not address specifically the pattern separation, although hyperexcitability of the GCs would decrease sparsity of the GC activity, leading to increase in overlap and then
to decrease in pattern separation.

To bridge the gap between the abstract computational models and the large-scale biological networks, a relatively small-scale simplified network, based on the prior ideas of the abstract models and including not only the GCs, but also the hilar cells, was developed to investigate the pattern separation \cite{Myers1,Myers2}. Recently, the effects of GC hyperexcitability and GC dendrites on the pattern separation have also been investigated in the  biological spiking neural networks \cite{Chavlis,Yim}. Here, we develop our spiking neural network for the hippocampal DG by following our prior approach for the cerebellar ring network \cite{Kim1,Kim2}. Our DG network is also based on the anatomical and the physiological properties described in \cite{Chavlis}.

\subsection{Architecture of The Spiking Neural Network of Hippocampal Dentate Gyrus}
\label{subsec:ABN}
Figure \ref{fig:DGN}(a) shows the box diagram for the hippocampal DG network. In the DG, we consider the granular layer, consisting of the excitatory GCs and the inhibitory BCs, and the underlying hilus, composed of the excitatory MCs and the inhibitory HIPP cells. We note that there are two types of excitatory cells, GCs and MCs, in contrast to the case of the CA3 and CA1 with only one type of excitatory pyramidal cells.

From the outside of the DG, the EC provides the external excitatory inputs into the GCs and the HIPP cells (that give inhibition to the GCs) via PPs. Thus, the GCs receive direct excitatory EC input via PPs and indirect disynaptic inhibitory EC input, mediated by the hilar HIPP cells.
The GCs are grouped into lamellar clusters \cite{Cluster1,Cluster2,Cluster3,Cluster4}, and one inhibitory BC exists in each GC cluster. Thus, a dynamical GC-BC loop is formed, and the BC (receiving the excitation from all the GCs) provides the feedback inhibition to the GCs.

The hilar MCs receive the excitation from the GCs, and provide feedback excitation to the GCs (i.e., formation of the GC-MC loop).
They control the activity of the GC-BC loop in each GC cluster by providing excitation to the GCs and the BC.
Thus, the GCs receive the direct excitatory MC input and the indirect disynaptic inhibitory MC input, mediated by the BC.
In this way, from the outside of the GC cluster, the GCs receive two types of excitatory inputs from the EC and the MCs and one kind of inhibitory input from the HIPP cells, and within the GC cluster they receive the feedback inhibition from the BC (receiving the excitation from the GCs and the MCs).

We follow our prior approach in the cerebellar ring network \cite{Kim1,Kim2}, develop a one-dimensional ring network for the hippocampal DG which has advantage for computational and analytical efficiency, and its visual representation may also be easily made, as in the famous small-world ring network \cite{SWN1,SWN2}.
As in \citep{Chavlis}, based on the anatomical data, we choose the numbers of the constituent cells (GCs, BCs, MCs, and HIPP cells) and the EC cells in our DG network and the connection probabilities between them.

Here, we consider a scaled-down spiking neural network where the total number of excitatory GCs is $N_{\rm GC}~(=2000)$, corresponding to $\frac {1}{500}$ of the $10^6$ GCs found in rats \cite{ANA1}. These GCs are grouped into the $N_c~(=100)$ lamellar clusters \cite{Cluster1,Cluster2,Cluster3,Cluster4}. In each GC cluster, there exist $n_{\rm {GC}}^{(c)}~(=20)$ GCs and one inhibitory BC. Thus, the number of BCs in the whole DG network becomes $N_{\rm BC}~(= 100)$, corresponding to $\frac {1} {20}$ of $N_{\rm GC}$.

In addition to the GCs and the BCs in the granular layer, the hilus consists of the excitatory MCs and the inhibitory HIPP cells.
In rats, the number of MCs, $N_{\rm MC},$ varies from 30,000 to 50,000, which correspond to 3-5 MCs per 100 GCs \cite{ANA2}.
Hence, we choose $N_{\rm MC}~=80$ in our DG network. Also, the estimated number of HIPP cells, $N_{\rm HIPP},$ is about 12,000, corresponding to
2 HIPP cells per 100 GCs. In our DG network, the number of the HIPP cells is $N_{\rm HIPP}=40.$

The EC layer II is the external source providing the excitatory inputs to the GCs and the HIPP cells via the PPs. The estimated number of the EC layer II cells, $N_{\rm EC},$ is about 200,000 in rats, which corresponds to 20 EC cells per 100 GCs  \cite{ANA3}. Hence, $N_{\rm EC}=400$ in our network.

Figure \ref{fig:DGN}(b1) shows a schematic diagram for the EC ring network, consisting of $N_{\rm EC}$ EC cells (denoted by the solid circles).
The activation degree $D_a$ of the EC cells is chosen as 10 $\%$ \cite{ANA4}. We randomly choose 40 active ones among the 400 EC (layer II) cells.
Each active EC cell is modeled in terms of the Poisson spike train with frequency of 40 Hz \cite{ANA5}. Here, the random-connection probability
$p^{\rm (GC,EC)}$ ($p^{\rm (HIPP,EC)}$) from the pre-synaptic EC cells to a post-synaptic GC (HIPP cell) is 20 $\%$. Hence, each GC or HIPP cell is randomly connected with the average number of 80 EC cells (among which the average number of active EC cells is just 8).

Figure \ref{fig:DGN}(b2) shows a schematic diagram for the granular-layer ring network with concentric inner GC and outer BC rings.
Numbers represent GC clusters (bounded by dotted lines). Each GC cluster ($I~=~1, \dots, N_c$) consists of GCs (solid circles) and one BC (diamond).
In our network, $N_c$ (number of the GC clusters) = 100 and $n_{\rm {GC}}^{(c)}$ (number of the GCs in each GC cluster) = 20. In each GC cluster, all the GCs provide excitation to the BC which then gives the feedback inhibition to all the GCs; there are no synaptic couplings between the GCs. Also, there are no intercluster interactions for the GCs and the BCs. In this way, the GCs and the BC forms an E-I dynamical loop in each GC cluster.

Figure \ref{fig:DGN}(b3) shows a schematic diagram for the hilar ring network with concentric inner MC and outer HIPP rings.
In our network, there are $N_{\rm MC}~(=80)$ MCs and $N_{\rm HIPP}~(=40)$ HIPP cells. Here, the MCs and the GCs are mutually connected with 20 $\%$ random-connection probabilities $p^{\rm (MC,GC)}$ ($\rm GC \rightarrow MC$) and $p^{\rm (GC,MC)}$ ($\rm MC \rightarrow GC$). In this way, the GCs and the MCs form a dynamical E-E loop.  All the MCs also provide the excitation to the BC in each GC cluster; the BC in the GC cluster receives excitatory inputs from all the
GCs in the same GC cluster and from all the MCs. In this way, the MCs control the activity of the GC-BC loop by providing excitation to the GCs and the BC in each GC cluster. Next, each GC in the GC cluster receives inhibition from the randomly-connected HIPP cells with the connection probability $p^{\rm (GC,HIPP)}~=~20 ~\%$. Then, the firing activity of the GCs is determined via competition between the excitatory inputs from the EC cells and from the MCs and the inhibitory input from the HIPP cells.

\subsection{Leaky Integrate-And-Fire Spiking Neuron Model with Afterhyperpolarization Current}
\label{subsec:LIF}
As elements of the hippocampal DG ring network, we choose leaky integrate-and-fire (LIF) spiking neuron models with additional afterhyperpolarization (AHP) currents,  determining refractory periods, as in our prior study of cerebellar ring network \cite{Kim1,Kim2}. This LIF spiking neuron model is one of the simplest spiking neuron models \cite{LIF}. Because of its simplicity, it can be easily analyzed and simulated. Hence, it has been very popularly used as a spiking neuron model.

The following equations govern evolution of dynamical states of individual cells in the $X$ population:
\begin{equation}
C_{X} \frac{dv_{i}^{(X)}}{dt} = -I_{L,i}^{(X)} - I_{AHP,i}^{(X)} + I_{ext}^{(X)} - I_{syn,i}^{(X)}, \;\;\; i=1, \cdots, N_{X},
\label{eq:GE}
\end{equation}
where $N_X$ is the total number of neurons in the $X$ population, $X=$ GC and BC in the granular layer and $X=$ MC and HIPP in the hilus.
In Eq.~(\ref{eq:GE}), $C_{X}$ (pF) denotes the membrane capacitance of the cells in the $X$ population, and the state of the $i$th cell in the $X$ population at a time $t$ (msec) is characterized by its membrane potential $v_i^{(X)}(t)$ (mV). The time-evolution of $v_i^{(X)}(t)$ is governed by 4 types of currents (pA) into the
$i$th cell in the $X$ population; the leakage current $I_{L,i}^{(X)}(t)$, the AHP current $I_{AHP,i}^{(X)}(t)$, the external constant current $I_{ext,i}^{(X)}$, and the synaptic current $I_{syn,i}^{(X)}(t)$. Here, we consider a subthreshold case of $I_{ext}^{(X)}=0$  for all $X$ \cite{Chavlis}.

In Eq.~(\ref{eq:GE}), the 1st type of leakage current $I_{L,i}^{(X)}$ for the $i$th neuron in the $X$ population is given by:
\begin{equation}
I_{L,i}^{(X)}(t) = g_{L}^{(X)} (v_{i}^{(X)}(t) - V_{L}^{(X)}),
\label{eq:Leakage}
\end{equation}
where $g_L^{(X)}$ and $V_L^{(X)}$ are conductance (nS) and reversal potential for the leakage current, respectively.
When its membrane potential $v_i^{(X)}$ reaches a threshold $v_{th}^{(X)}$ at a time $t_{f,i}^{(X)}$, the $i$th neuron in the $X$ population
fires a spike. After spiking (i.e., $t \geq t_{f,i}^{(X)}$), the 2nd type of AHP current $I_{AHP,i}^{(X)}$ follows:
\begin{equation}
I_{AHP,i}^{(X)}(t) = g_{AHP}^{(X)}(t) ~(v_{i}^{(X)}(t) - V_{AHP}^{(X)})~~~{\rm ~for~} \; t \ge t_{f,i}^{(X)}.
\label{eq:AHP1}
\end{equation}
Here, $V_{AHP}^{(X)}$ is the reversal potential for the AHP current, and the conductance $g_{AHP}^{(X)}(t)$ is given by an exponential-decay
function:
\begin{equation}
g_{AHP}^{(X)}(t) = \bar{g}_{AHP}^{(X)}~  e^{-(t-t_{f,i}^{(X)})/\tau_{AHP}^{(X)}} ,
\label{eq:AHP2}
\end{equation}
where $\bar{g}_{AHP}^{(X)}$ and $\tau_{AHP}^{(X)}$ are the maximum conductance and the decay time constant for the AHP current.
With increasing $\tau_{AHP}^{(X)}$, the refractory period becomes longer.

In Appendix \ref{app:A}, the parameter values for the capacitance $C_X$, the leakage current $I_L^{(X)}$, and the AHP current $I_{AHP}^{(X)}$ are shown in Table \ref{tab:Singleparm}. These values are based on physiological properties of the GC, BC, MC, and HIPP cell \cite{Chavlis,Hilus3}.

\subsection{Synaptic Currents}
\label{subsec:SC}
In Eq.~(\ref{eq:GE}). the synaptic current $I_{syn,i}^{(X)}$ into the $i$th neuron in the $X$ population consists of the following 3 kinds of synaptic currents:
\begin{equation}
I_{syn,i}^{(X)} = I_{{\rm AMPA},i}^{(X,Y)} + I_{{\rm NMDA},i}^{(X,Y)} + I_{{\rm GABA},i}^{(X,Z)}.
\label{eq:ISyn1}
\end{equation}
Here, $I_{{\rm AMPA},i}^{(X,Y)}$ and $I_{{\rm NMDA},i}^{(X,Y)}$ are the excitatory AMPA ($\alpha$-amino-3-hydroxy-5-methyl-4-isoxazolepropionic acid) receptor-mediated and NMDA ($N$-methyl-$D$-aspartate) receptor-mediated currents from the pre-synaptic source $Y$ population to the post-synaptic $i$th neuron in the target $X$ population. In contrast, $I_{{\rm GABA},i}^{(X,Z)}$ is the inhibitory $\rm GABA_A$ ($\gamma$-aminobutyric acid type A) receptor-mediated current
from the pre-synaptic source $Z$ population to the post-synaptic $i$th neuron in the target $X$ population.

As in the case of the AHP current, the $R$ (= AMPA, NMDA, or GABA) receptor-mediated synaptic current $I_{R,i}^{(T,S)}$ from the pre-synaptic source $S$ population to the $i$th post-synaptic neuron in the target $T$ population is given by:
\begin{equation}
I_{R,i}^{(T,S)}(t) = g_{R,i}^{(T,S)}(t)~(v_{i}^{(T)}(t) - V_{R}^{(S)}),
\label{eq:ISyn2}
\end{equation}
where $g_{(R,i)}^{(T,S)}(t)$ and $V_R^{(S)}$ are synaptic conductance and synaptic reversal potential
(determined by the type of the pre-synaptic source $S$ population), respectively.
We obtain the synaptic conductance $g_{R,i}^{(T,S)}(t)$ from:
\begin{equation}
g_{R,i}^{(T,S)}(t) = K_{R}^{(T,S)} \sum_{j=1}^{N_S} w_{ij}^{(T,S)} ~ s_{j}^{(T,S)}(t),
\label{eq:ISyn3}
\end{equation}
where $K_{R}^{(T,S)}$ is the synaptic strength per synapse for the $R$-mediated synaptic current
from a pre-synaptic neuron in the source $S$ population to a post-synaptic neuron in the target $T$ population.
The inter-population synaptic connection from the source $S$ population (with $N_s$ neurons) to the target $T$ population is given by the connection weight matrix
$W^{(T,S)}$ ($=\{ w_{ij}^{(T,S)} \}$) where $w_{ij}^{(T,S)}=1$ if the $j$th neuron in the source $S$ population is pre-synaptic to the $i$th neuron
in the target $T$ population; otherwise $w_{ij}^{(T,S)}=0$.

The post-synaptic ion channels are opened due to the binding of neurotransmitters (emitted from the source $S$ population) to receptors in the target
$T$ population. The fraction of open ion channels at time $t$ is denoted by $s^{(T,S)}$. The time course of $s_j^{(T,S)}(t)$ of the $j$th neuron
in the source $S$ population is given by a sum of double exponential functions $E_{R}^{(T,S)} (t - t_{f}^{(j)}-\tau_{R,l}^{(T,S)})$:
\begin{equation}
s_{j}^{(T,S)}(t) = \sum_{f=1}^{F_{j}^{(S)}} E_{R}^{(T,S)} (t - t_{f}^{(j)}-\tau_{R,l}^{(T,S)}),
\label{eq:ISyn4}
\end{equation}
where $t_f^{(j)}$ and $F_j^{(S)}$ are the $f$th spike time and the total number of spikes of the $j$th neuron in the source $S$ population, respectively.
$\tau_{R,l}^{(T,S)}$ is the synaptic latency time constant for $R$-mediated synaptic current.
The exponential-decay function $E_{R}^{(T,S)} (t)$ (which corresponds to contribution of a pre-synaptic spike occurring at $t=0$ in the absence of synaptic latency)
is given by:
\begin{equation}
E_{R}^{(T,S)}(t) = \frac{1}{\tau_{R,d}^{(T,S)}-\tau_{R,r}^{(T,S)}} \left( e^{-t/\tau_{R,d}^{(T,S)}} - e^{-t/\tau_{R,r}^{(T,S)}} \right) \cdot \Theta(t), \label{eq:ISyn5}
\end{equation}
where $\Theta(t)$ is the Heaviside step function: $\Theta(t)=1$ for $t \geq 0$ and 0 for $t <0$. $\tau_{R,r}^{(T,S)}$ and $\tau_{R,d}^{(T,S)}$ are synaptic rising and decay time constants of the $R$-mediated synaptic current, respectively.

In Appendix \ref{app:A}, Tables \ref{tab:Synparm1} and \ref{tab:Synparm2} show the parameter values for the synaptic strength per synapse $K_{R}^{(T,S)}$,
the synaptic rising time constant $\tau_{R,r}^{(T,S)}$, synaptic decay time constant $\tau_{R,d}^{(T,S)}$, synaptic latency time constant $\tau_{R,l}^{(T,S)}$, and the synaptic reversal potential  $V_{R}^{(S)}$ for the synaptic currents into the GCs and for the synaptic currents into the HIPP cells, the MCs and the BCs, respectively. These parameter values are also based on the physiological properties of the relevant cells \cite{Chavlis,SynParm1,SynParm2,SynParm3,SynParm4,SynParm5,SynParm6,SynParm7,SynParm8}.

Numerical integration of the governing Eq.~(\ref{eq:GE}) for the time-evolution of states of individual spiking neurons
is done by employing the 2nd-order Runge-Kutta method with the time step 0.1 msec.
We choose random initial points $v_i^{(X)}(0)$ for the $i$th neuron in the $X$ population
with uniform probability in the range of $v_i^{(X)}(0) \in (V_L^{(X)}-5.0, V_L^{(X)}+5.0)$; the values of $V_L^{(X)}$ are
given in Table \ref{tab:Singleparm}.

\begin{figure}[t]
\includegraphics[width=\columnwidth]{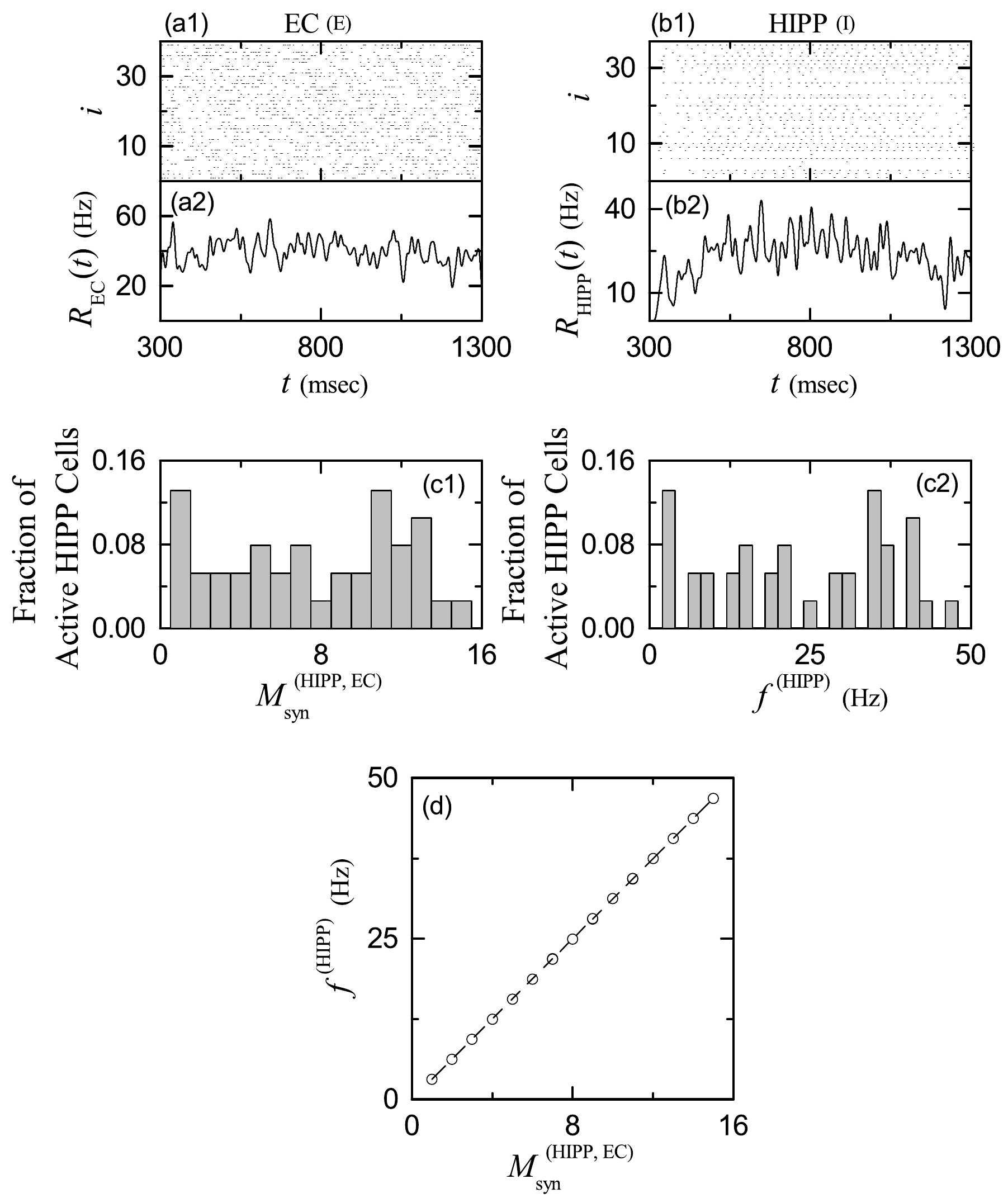}
\caption{External inputs from the EC. Direct excitatory (E) EC input via PP: (a1) Raster plot of spikes of 40 active EC cells. (a2) Instantaneous population spike rate $R_{\rm EC}(t)$ of active EC cells. Band width for $R_{\rm EC}(t)$: $h=20$ msec. Indirect disynaptic inhibitory (I) EC input, mediated by the hilar HIPP cells: (b1) Raster plot of spikes of 37 active HIPP cells. (b2) Instantaneous population spike rate $R_{\rm HIPP}(t)$ of active HIPP cells. Band width for $R_{\rm HIPP}(t)$: $h=20$ msec. (c1) Histogram of $M_{\rm {syn}}^{\rm {(HIPP,EC)}}$ (number of pre-synaptic active EC cells onto the post-synaptic active HIPP cells). Bin size for the histogram is 1. (c2) Histogram of $f^{\rm {(HIPP)}}$ [mean firing rate (MFR) of active HIPP cells]. Bin size for the histogram is 2 Hz. (d) Plot of $f^{\rm {(HIPP)}}$ versus $M_{\rm {syn}}^{\rm {(HIPP,EC)}}$.
}
\label{fig:EC}
\end{figure}

\section{Dynamical Origin for The Winner-Take-All Competition}
\label{sec:DO}
As a pre-processor for the CA3, the GCs in the DG perform the pattern separation, facilitating the pattern storage and completion in the CA3.
The GCs exhibit sparse firing activity through competitive learning, which has been thought to improve the pattern separation.
In this section, we investigate the dynamical origin of the WTA competition, leading to the sparse activation of the GCs.
The firing activity of the GCs may be well determined in terms of the E-I conductance ratio ${ {\cal{R}}_{\rm E-I}^{\rm (con)}}^*$
(given by the time average of the ratio of the external E to I conductances). GCs become active (i.e., they become winners) only when their
${ {\cal{R}}_{\rm E-I}^{\rm (con)}}^*$ is larger than a threshold ${\cal R}_{th}^*$. WTA competition is thus found to occur via
competition between the firing activity of the GCs and the feedback inhibition of the BC in each GC cluster. In this case, the hilar MCs
is also found to play an essential role to enhance the WTA competition by providing excitation to both the GCs and the BC.

\subsection{Firing Activity of GCs in The Presence of The External Excitatory EC and The Inhibitory HIPP Inputs}
\label{subsec:GCFR}
In this subsection, we study firing activity of the GCs under the external excitatory input from the EC cells and the inhibitory input
from the HIPP cells. The firing activity of the GCs is found to be determined via competition between the excitatory EC input and the inhibitory
HIPP input. Particularly, such competition  may be well represented in terms of the E-I synapse ratio ${\cal R}_{\rm E-I}^{\rm (syn)},$ given by the
ratio of the number of the pre-synaptic EC cells ($M_{\rm syn}^{\rm (GC,EC)}$) to the number of the pre-synaptic HIPP cells ($M_{\rm syn}^{\rm (GC,HIPP)}$).

Figure \ref{fig:EC} shows the external input from the EC. There are direct excitatory input from the EC cells and indirect disynaptic inhibitory EC input, mediated by the hilar HIPP cells [see Fig.~\ref{fig:DGN}(a)]. Among the 400 EC cells, randomly-chosen 40 active cells make spikings (i.e., activation degree $D_a=10$ $\%$). Each active EC cell is modeled in terms of the Poisson spike train with frequency of 40 Hz. After a break stage ($t=0-300$ msec), Poisson spike train of each active EC cell follows during the stimulus stage ($t=300- 30,300$ msec; the stimulus period $T_s$ is $3 \cdot 10^4$ msec). Then, population firing activity of the active EC cells may be well visualized in the raster plot of spikes in Fig.~\ref{fig:EC}(a1) which is a collection of spike trains of individual active EC cells; for convenience, only a part from $t=300$ to 1,300 msec is shown in the raster plot of spikes. Spikes of the active EC cells are completely scattered without forming any synchronized ``spiking stripes,'' and hence the population state of the active ECs becomes desynchronized.

As a population quantity showing collective firing behaviors, we use an instantaneous population spike rate (IPSR) which may be obtained from the raster plots of spikes \cite{RM,W_Review,Sparse1,Sparse2,Sparse3,FSS}. To get a smooth IPSR, we employ the kernel density estimation (kernel smoother) \cite{Kernel}. Each spike in the raster plot is convoluted (or blurred) with a kernel function $K_h(t)$ to get a smooth estimate of IPSR $R_{\rm EC}(t)$:
\begin{equation}
R_{\rm{EC}}(t) = \frac{1}{N_a} \sum_{i=1}^{N_a} \sum_{s=1}^{n_i} K_h (t-t_{s}^{(i)}),
\label{eq:IPSR}
\end{equation}
where $N_a$ is the number of the active cells, $t_{s}^{(i)}$ is the $s$th spiking time of the $i$th active cell, $n_i$ is the total number of spikes for the $i$th active cell, and we use a Gaussian kernel function of band width $h$:
\begin{equation}
K_h (t) = \frac{1}{\sqrt{2\pi}h} e^{-t^2 / 2h^2}, ~~~~ -\infty < t < \infty.
\label{eq:Gaussian}
\end{equation}
Throughout the paper, the band width $h$ of $K_h(t)$ is 20 msec. The IPSR $R_{\rm {EC}}(t)$ of the active EC cells is shown in Fig.~\ref{fig:EC}(a2), and it shows
relatively small noisy fluctuations around its time average (i.e., $\overline{R_{\rm {EC}}(t)} =40$ Hz) without distinct synchronous oscillations.

The active EC cells provide direct excitatory input and indirect disynaptic inhibitory input, mediated by the HIPP cells, to the GCs. Thus, the EC cells and the
HIPP cells become the excitatory and the inhibitory input sources to the GCs, respectively. We note that each HIPP cell is randomly connected to the average number of 80 EC cells with the connection probability $p^{\rm (HIPP,EC)}$ = 20$\%$, among which the average number of active EC cells is 8. Figures \ref{fig:EC}(b1) and \ref{fig:EC}(b2) show the raster plot of spikes of the active HIPP cells and the corresponding IPSR $R_{\rm{HIPP}}(t)$. Among the 40 HIPP cells, 37 HIPP cells are active, while the remaining 3 HIPP cells (without receiving excitatory input from the active EC cells) are silent; the activation degree of the HIPP cells is 92.5$\%$. Also, the spiking of the active HIPP cells begins from $t \simeq 320$ msec [i.e. about 20 msec delay for the firing onset of the HIPP cells with respect to
the firing onset ($t=300$ msec) of the active EC cells].

\begin{figure}[t]
\includegraphics[width=0.7\columnwidth]{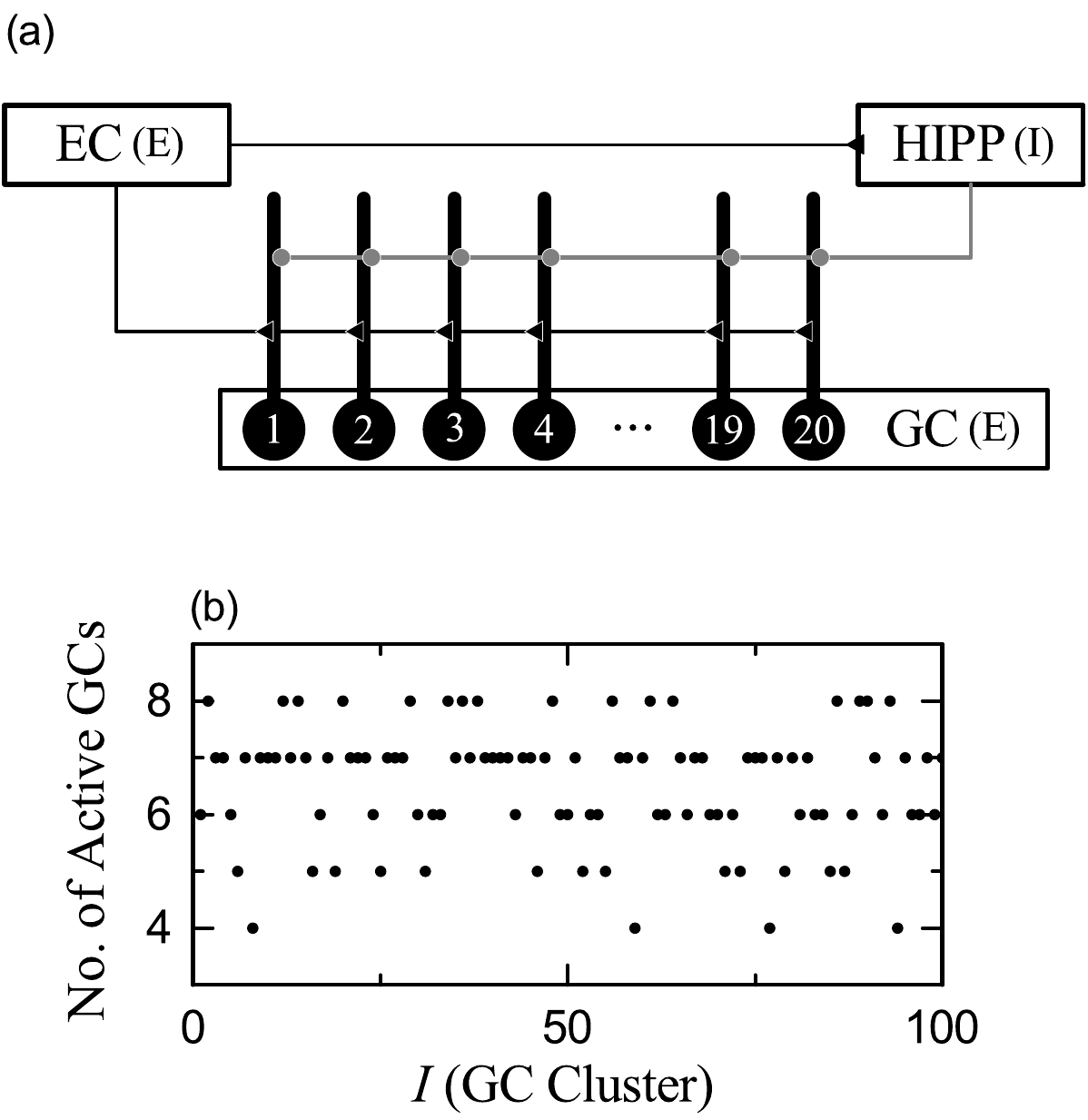}
\caption{Firing activity of GCs in the presence of only the external direct excitatory EC input and indirect disynaptic inhibitory EC input, mediated by the HIPP cells. (a) Diagram for the external direct excitatory input from the EC cells (black line with triangles) and indirect disynaptic inhibitory inputs from the EC cells mediated by the HIPP cells (gray line with circles) into the GCs. (b) Plot of number of active GCs versus $I$ (GC cluster).
}
\label{fig:EC-GC}
\end{figure}

As in the case of the active EC cells, no synchronized spiking stripes are shown in the raster plot of spikes of the active HIPP cells.
However, unlike the case of the active EC cells (showing stochastic firing activity), the spike train of each active HIPP cell seems to be quasi-regular with its own MFR (i.e., each active HIPP cell seems to exhibit a quasi-regular firing activity). However, their MFRs seem to vary very differently depending on the active HIPP cells. Due to such diverse MFRs, no synchronized spiking stripes appear in the raster plot of spikes of the active HIPP cells. Hence, their IPSR $R_{\rm{HIPP}}(t)$ also shows noisy fluctuations around its time average (i.e., $\overline{R_{\rm {HIPP}}(t)} = 23$ Hz) without synchronous oscillations.

In Figs.~\ref{fig:EC}(c1)-\ref{fig:EC}(c2), we discuss how MFRs of the active HIPP cells become diverse.
The number of pre-synaptic active EC cells $M_{\rm syn}^{\rm (HIPP,EC)}$ for the post-synaptic active HIPP cells is broadly distributed in Fig.~\ref{fig:EC}(c1).
Its range is [1, 15], the mean is 7.8, and the standard deviation from the mean is 4.5; 3 silent HIPP cells have no active
pre-synaptic EC cells. The MFR $f^{\rm {(HIPP)}}$ of each active HIPP cell is obtained by dividing the total number of spikes by the stimulus
period $T_s$ $(=3\cdot 10^4$ msec). Figure \ref{fig:EC}(c2) shows broad distribution of the MFRs.
Its range is [2.6, 47.8] Hz, the population-averaged MFR $\langle f^{\rm {(HIPP)}} \rangle =22.9$ Hz, and the standard deviation from
$\langle f^{\rm {(HIPP)}} \rangle$ is 14.1 Hz. Because of these diverse MFRs, the active HIPP cells exhibit no collective synchronized
firing activity. We also note that there exists a strong positive correlation (with the Pearson's correlation coefficient $r=0.9999$)
between $M_{\rm syn}^{\rm (HIPP,EC)}$ (number of pre-synaptic active EC cells) and $f^{\rm {(HIPP)}}$  (MFRs of the post-synaptic HIPP cells)
[see Fig.~\ref{fig:EC}(d)] \cite{Pearson}; the larger the number of pre-synaptic active EC cells, the higher the MFR of the (post-synaptic) active HIPP cell.

\begin{figure}
\includegraphics[width=0.9\columnwidth]{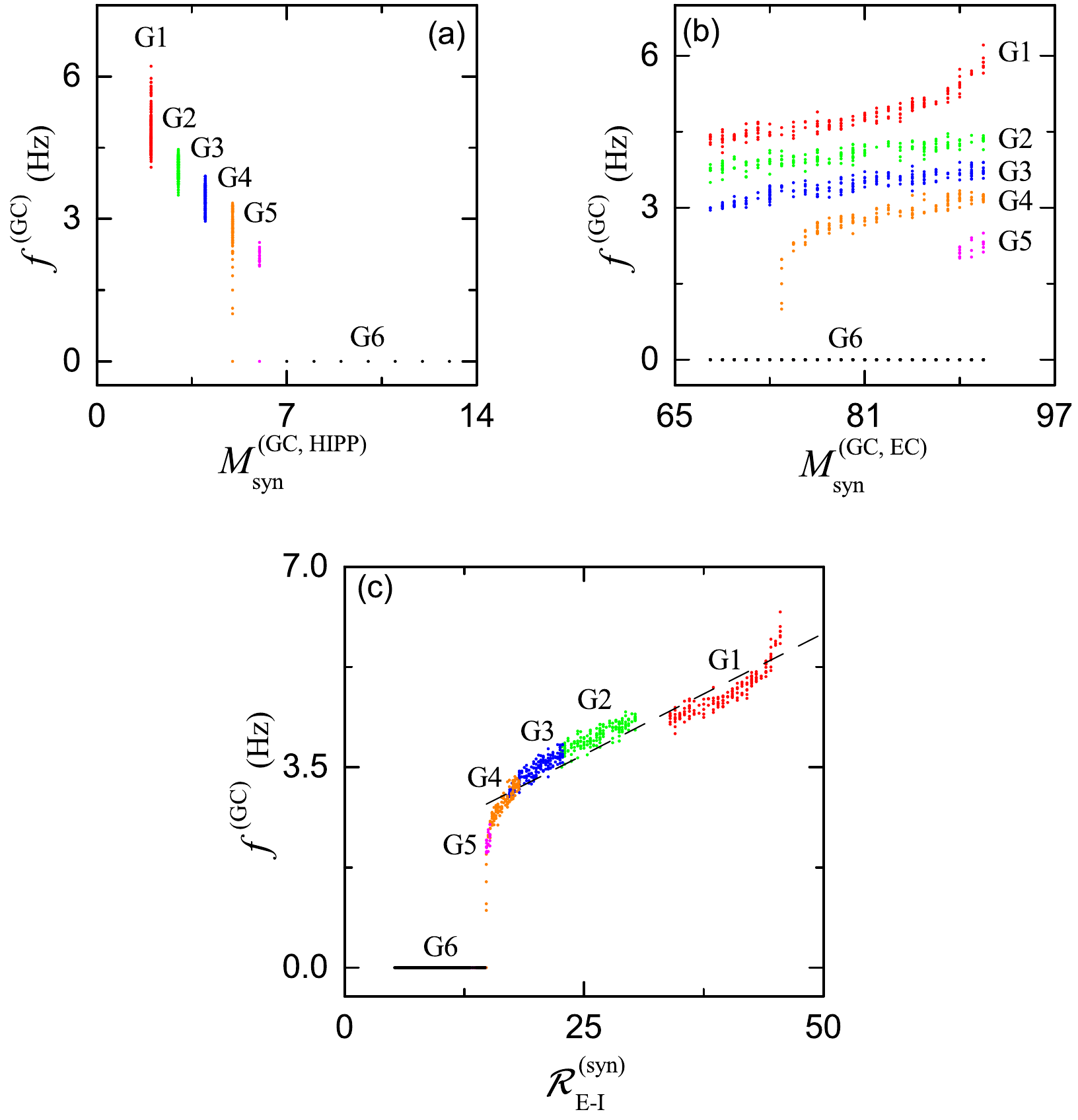}
\caption{Firing activity of GCs via competition between the numbers of pre-synaptic EC and HIPP cells. Plots of $f^{\rm {(GC)}}$ (MFR of GCs) versus (a)
$M_{\rm syn}^{\rm (GC,HIPP)}$ (number of pre-synaptic HIPP cells)  and (b) $M_{\rm syn}^{\rm (GC,EC)}$ (number of pre-synaptic EC cells).
GCs are decomposed into groups Gn $(n=1,\dots,6)$ with different number of pre-synaptic HIPP cells $M_{\rm syn}^{\rm (GC,HIPP)}$: G1 (red color online) [$M_{\rm syn}^{\rm (GC,HIPP)}=2$], G2 (green) [$M_{\rm syn}^{\rm (GC,HIPP)}=3$], G3 (blue) [$M_{\rm syn}^{\rm (GC,HIPP)}=4$], G4 (orange) [$M_{\rm syn}^{\rm (GC,HIPP)}=5$], G5 (violet) [$M_{\rm syn}^{\rm (GC,HIPP)}=6$], and G6 (black) [$M_{\rm syn}^{\rm (GC,HIPP)} \geq 7$].
(c) Plot of $f^{\rm {(GC)}}$ (MFR of GCs) versus the E-I synapse ratio ${\cal R}_{\rm E-I}^{\rm (syn)}$ $(= M_{\rm syn}^{\rm (GC,EC)}~/~ M_{\rm syn}^{\rm (GC,HIPP)})$; a fitted dashed line is given.
}
\label{fig:E-ICom}
\end{figure}

Then, we first investigate firing activity of GCs in the presence of only the external direct excitatory EC input and indirect disynaptic inhibitory EC input, mediated by the HIPP cells. Figure \ref{fig:EC-GC}(a) shows a diagram for the external direct excitatory input from the EC cells (black line with triangles) and indirect disynaptic inhibitory inputs from the EC cells, mediated by the HIPP cells (gray line with circles) into the GCs. In this case, the number of active GCs
in each GC cluster ($I=1, \dots, 100$) is shown in Fig.~\ref{fig:EC-GC}(b). A GC with at least one spike during the stimulus period $T_s$ ($=3 \cdot 10^4$ msec)
is active; otherwise, silent. In this case, the total number of active GCs is 652, and hence the activation degree is $D_a=32.6 \%.$
For the distribution of the number of active GCs in each GC cluster, its range is [4, 8], the mean is 6.52, and the standard deviation from the mean is
1.04.

In the case of Fig.~\ref{fig:EC-GC}(a), firing activity of the GCs is determined via competition between the direct excitatory EC input and the indirect disynaptic
inhibitory EC input, mediated by the HIPP Cells. The strength of direct excitatory EC input may be represented by the number of pre-synaptic EC cells,
$M_{\rm syn}^{\rm (GC,EC)}$ and the strength of indirect inhibitory EC input, mediated by the HIPP cells, can also be denoted by the number of pre-synaptic
HIPP cells, $M_{\rm syn}^{\rm (GC,HIPP)}.$ Then, the E-I synapse ratio ${\cal R}_{\rm E-I}^{\rm (syn)}$, defined by:
\begin{equation}
\label{eq:Rsyn}
 {\cal R}_{\rm E-I}^{\rm (syn)} = \frac {M_{\rm syn}^{\rm (GC,EC)}} {M_{\rm syn}^{\rm (GC,HIPP)}},
\end{equation}
represents well the competition between the excitatory input from the EC cells and the inhibitory input from the HIPP cells.

Figure \ref{fig:E-ICom}(a) shows $f^{\rm {(GC)}}$  (MFR  of the GCs) versus $M_{\rm {syn}}^{\rm {(GC,HIPP)}}$ (number of the pre-synaptic HIPP cells).
For the distribution of $M_{\rm {syn}}^{\rm {(GC,HIPP)}}$, its range is [2, 13], the mean is 7.9, and the standard deviation from the mean is 3.5.
Depending on $M_{\rm {syn}}^{\rm {(GC,HIPP)}}$, the whole GCs are decomposed into the 6 groups $\rm {Gn}$ ($\rm {n}=1, \dots,6)$ with different values of
$M_{\rm {syn}}^{\rm {(GC,HIPP)}}$. In the group G1 (red color online) with $M_{\rm {syn}}^{\rm {(GC,HIPP)}}=2,$ G2 (green) with $M_{\rm {syn}}^{\rm {(GC,HIPP)}}=3,$ G3 (blue) with $M_{\rm {syn}}^{\rm {(GC,HIPP)}}=4,$ G4 (orange) with $M_{\rm {syn}}^{\rm {(GC,HIPP)}}=5,$ G5 (violet) with $M_{\rm {syn}}^{\rm {(GC,HIPP)}}=6,$  and
G6 (black) with $M_{\rm {syn}}^{\rm {(GC,HIPP)}} \geq 7,$ the number of GCs (fraction) is 156 (7.8$\%$), 169 (8.45$\%$), 173 (8.65$\%$), 170 (8.50$\%$), 152 (7.6$\%$), and 1180 (59$\%$), respectively, With increasing $M_{\rm {syn}}^{\rm {(GC,HIPP)}},$ MFRs $f^{\rm {(GC)}}$  of the GCs tend to decrease due to increase
in the inhibitory input from the HIPP cells. In the groups G1, G2, and G3, only active GCs appear. On the other hand, from the group G4, silent GCs also appear along with active GCs, and eventually in the group G6, all the GCs are silent.

Figure \ref{fig:E-ICom}(b) shows $f^{\rm {(GC)}}$  (MFR  of the GCs) versus $M_{\rm {syn}}^{\rm {(GC,EC)}}$ (number of the pre-synaptic EC cells).
For the distribution of $M_{\rm {syn}}^{\rm {(GC,EC)}}$, its range is (68, 91), the mean is 79.4, and the standard deviation from the mean is 6.8.
In each Gn (n = 1, 2, 3) group, all the GCs are active, and their MFRs $f^{\rm {(GC)}}$ increase with $M_{\rm {syn}}^{\rm {(GC,EC)}}$ due to increase in excitation from the EC cells. On the other hand, in the G4 and G5 groups, when $M_{\rm {syn}}^{\rm {(GC,EC)}}$ passes a threshold $M_{th}^*$, active GCs begin to appear and then their MFRs $f^{\rm {(GC)}}$ also increase with $M_{\rm {syn}}^{\rm {(GC,EC)}}$; $M_{th}^*=$ 74 and 89 for G4 and G5, respectively. In the group G6 with $M_{\rm {syn}}^{\rm {(GC,HIPP)}} \geq 7,$ only silent GCs exist, independently of $M_{\rm {syn}}^{\rm {(GC,EC)}}.$

Figure \ref{fig:E-ICom}(c) shows $f^{\rm {(GC)}}$ (MFR  of the GCs) versus ${\cal R}_{\rm E-I}^{\rm (syn)}$ (E-I synapse ratio).
${\cal R}_{\rm E-I}^{\rm (syn)}$ denotes well the competition between the excitatory EC input and the inhibitory HIPP input.
We note that there exists a threshold ${\cal R}_{\rm {E-I}}^{\rm (syn),*}~(=14.8),$ above which active GCs appear.
With increasing ${\cal R}_{\rm {E-I}}^{\rm (syn)}$ from the threshold ${\cal R}_{\rm {E-I}}^{\rm (syn),*}$, MFRs $f^{\rm {(GC)}}$ increase monotonically.
In the active region, $f^{\rm {(GC)}}$ shows a strong correlation with ${\cal R}_{\rm {E-I}}^{\rm (syn)}$ with the Pearson's correlation coefficient
$r=0.9388$.

\begin{figure}
\includegraphics[width=0.7\columnwidth]{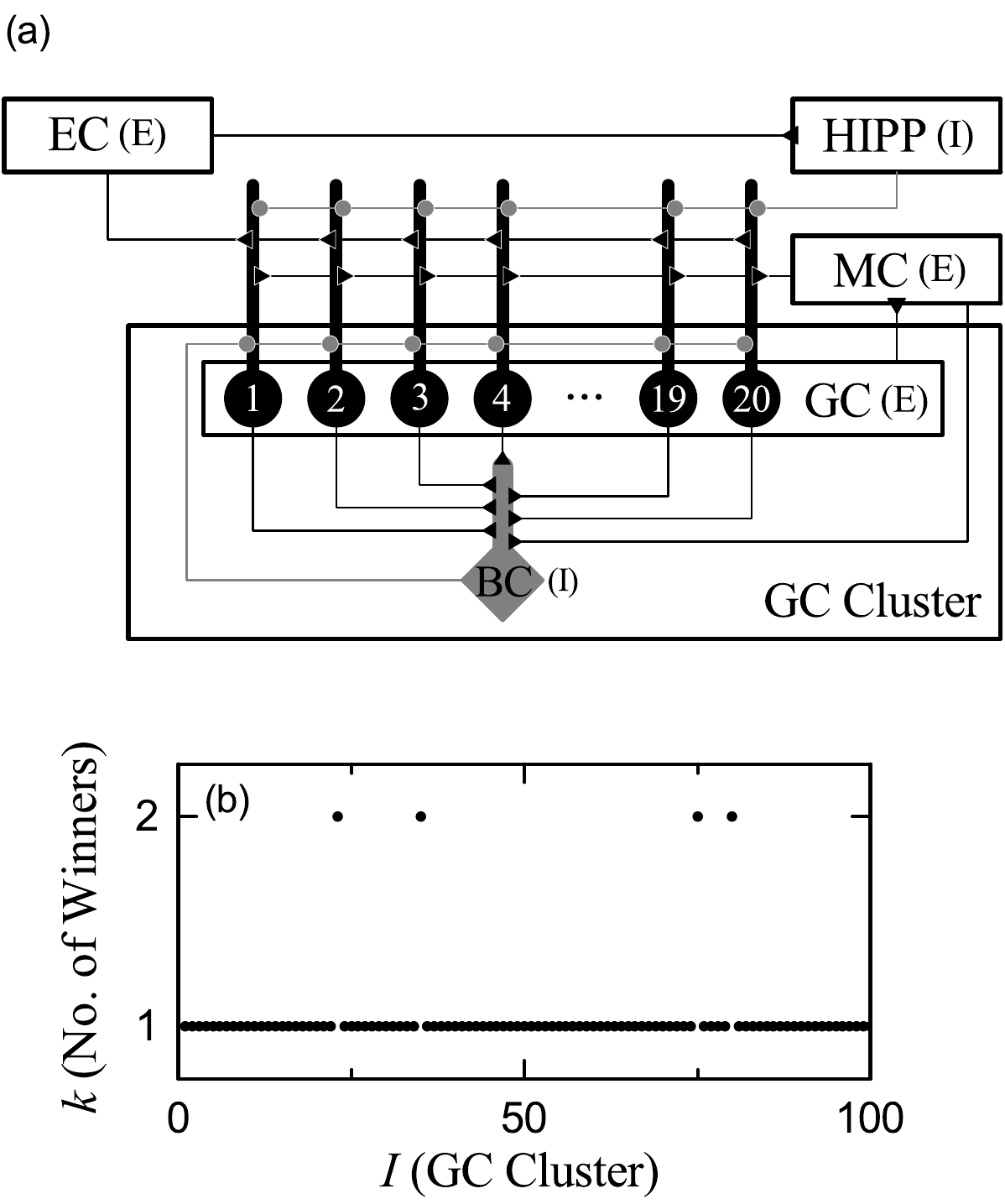}
\caption{WTA competition in the whole DG network. (a) Diagram of the whole DG network for a GC cluster, composed of 20 excitatory GCs and one inhibitory BC (basket cell). There are three kinds of external inputs: two types of excitatory inputs from the EC cells and the MCs (black line with triangles) and one kind of inhibitory inputs from the HIPP cells (gray line with solid circles). In the GC cluster, all GCs excite the BC which provides feedback inhibition to all the GCs. (b) Plot of $k$ (number of active GCs) versus $I$ (GC cluster index).
}
\label{fig:WN}
\end{figure}

\subsection{WTA Competition in The Whole DG Network}
\label{subsec:WN}
In this subsection, we investigate the WTA competition in the whole DG network, composed of the hilar MCs and the BCs, in addition to the EC cells, the HIPP cells, and the GCs in Fig.~\ref{fig:EC-GC}(a). Figure \ref{fig:WN}(a) shows a diagram of the whole DG network for a GC cluster,
consisting of 20 excitatory GCs and one inhibitory BC. There are three kinds of external inputs to the GCs: two types of excitatory inputs from the EC cells and the MCs  (black line with triangles) and one kind of inhibitory inputs from the HIPP cells (gray line with solid circles). Within the GC cluster, all GCs excite the BC which then provides feedback inhibition to all the GCs. In comparison to the case in Fig.~\ref{fig:EC-GC}(a), one more excitatory input from the MCs occurs.
These MCs tend to facilitate firing activity of the GC-BC loop in the GC cluster by providing the excitatory inputs to both the GCs and the BC.

\begin{figure}
\includegraphics[width=0.9\columnwidth]{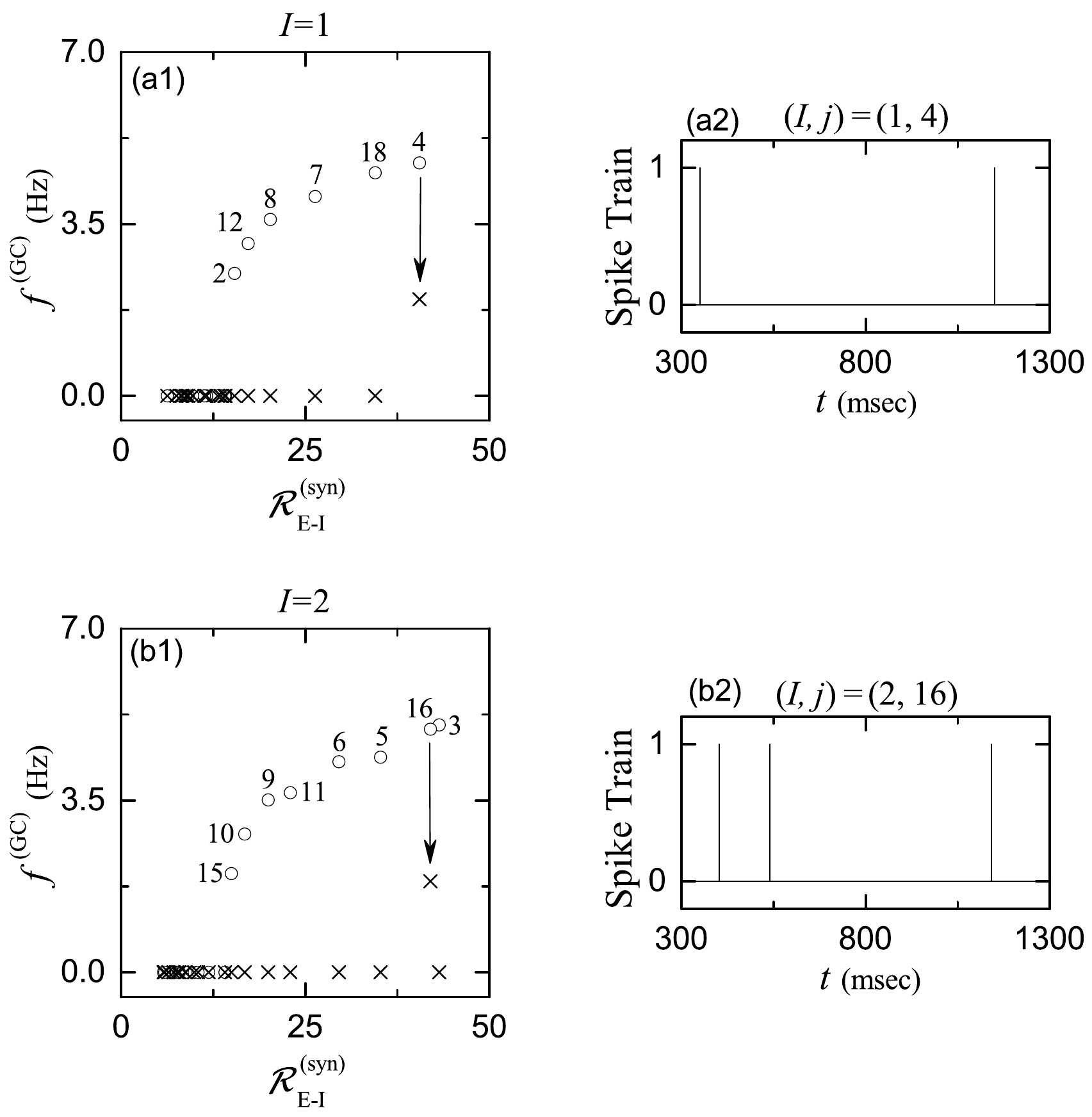}
\caption{$k = 1$ WTA competition. (a1) Plot of $f^{\rm (GC)}$ (MFR of GCs) versus ${\cal R}_{\rm E-I}^{\rm (syn)}$ [E-I synapse ratio in Eq.~(\ref{eq:Rsyn})]
in the 1st $(I = 1)$ GC cluster. (a2) Spike train of the winner GC $(j = 4)$ in the 1st GC cluster. (b1) Plot of $f^{\rm (GC)}$ (MFR of GCs) versus
${\cal R}_{\rm E-I}^{\rm (syn)}$ in the 2nd $(I = 2)$ GC cluster. (b2) Spike train of the winner GC $(j = 16)$ in the 2nd GC cluster. Open circles and crosses in (a1) and (b1) represent firing activity of GCs in the presence of only the external inputs from the EC and the HIPP cells [shown in Fig.~\ref{fig:EC-GC}(a)] and in the whole network [shown in Fig.~\ref{fig:WN}(a)], respectively. Numbers in (a1) and (b1) represent the GC index in each GC cluster.
}
\label{fig:k1WTA}
\end{figure}

\begin{figure}
\includegraphics[width=0.9\columnwidth]{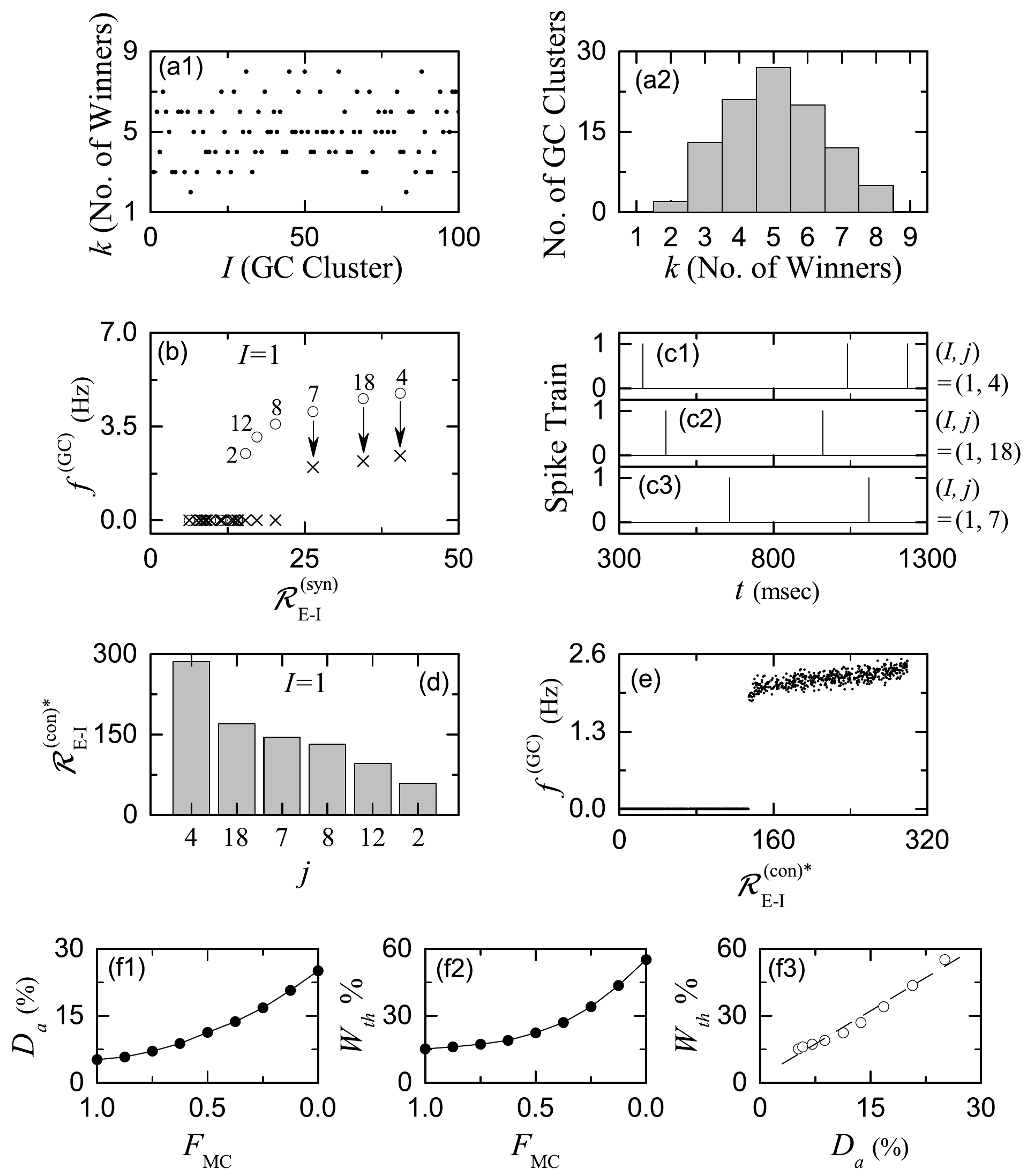}
\caption{WTA competition in the GC-BC loop in the case of $F_{\rm MC}$ (fraction of MCs) = 0 (i.e., complete loss of MCs) [(a1) - (e)]. (a1) Plot of
$k$ (number of winner GCs) versus $I$ (GC cluster index). (a2) Plot of the number of GC clusters versus $k$ (number of winners). 1st ($I = 1$) GC cluster in (b1)-(d). (b) Plot of $f^{\rm (GC)}$ (MFR of GCs) versus ${\cal R}_{\rm E-I}^{\rm (syn)}$ (E-I synapse ratio).
Open circles and crosses represent firing activity of GCs in the presence of only the external inputs from the EC and the HIPP cells and in the presence of
the additional feedback inhibition from the BC, respectively. (c1) - (c3) Spike trains of the three winner GCs ($j =$ 4, 18, and 7), and
(d) Plot of the E-I conductance ratio ${ {\cal{R}}_{\rm E-I}^{\rm (con)}}^*$ versus the 6 active GCs in the presence of only the excitatory EC input and the inhibitory HIPP input [$j$ ($x$-axis label) represents the GC index]. (e) Plot of $f^{\rm (GC) }$ (MFR of all the GCs) versus ${ {\cal{R}}_{\rm E-I}^{\rm (con)}}^*$.
(f1) Plot of $D_a$ (activation degree) of GCs versus $F_{\rm MC}$ (fraction of MCs). (f2) Plot of the winner threshold percentage $W_{th}$$\%$ versus $F_{\rm MC}$. (f3) Plot of  $W_{th}$$\%$ versus $D_a$ of GCs.
}
\label{fig:MC1}
\end{figure}

In the whole DG network, firing activity of the GCs is determined via competition between the two excitatory EC and MC inputs and the one inhibitory HIPP input.
Then, within the GC cluster, interaction of excitation of the GCs with feedback inhibition from the BC leads to WTA competition.
Figure \ref{fig:WN}(b) shows the plot of number of active GCs versus $I$ (GC cluster index).
Only one active winner ($k=1$) GC appears in most GC clusters, except for the 4 clusters ($I=$23, 35, 75, and 80) where $k=2$ winners exist;
96$\%$ GC clusters with $k=1$ winner and 4$\%$ GC clusters with $k=2$ winners.
Consequently, the total number of active GCs is 104, corresponding to $D_a=5.2\%$ (activation degree of the GCs).
In comparison to $D_a=32.6\%$ in the presence of only the excitatory EC input and the inhibitory HIPP input (Fig.~\ref{fig:EC-GC}),
sparse firing activity of the GCs results from the excitation from the MCs (facilitating firing activity of the GC-BC loop) and the feedback inhibition from the BC.
Without the MCs, the activation degree of the GCs becomes increased to $D_a=25.1 \%$ only due to the feedback inhibitory BC input, which will
be discussed in details in Fig.~\ref{fig:MC1}. Consequently, the MCs tend to enhance the WTA competition through facilitating firing activity of the GC-BC loop.

As an example, we first consider the $k=1$ WTA competition in the $I=1$ GC cluster.
Firing activity of the GCs is determined via competition between the external excitatory and inhibitory inputs into the GCs.
Then, feedback inhibition from the BC selects which GCs fire. Only strongly active GCs may survive under the feedback inhibition.
Figure \ref{fig:k1WTA}(a1) shows the plot of $f^{\rm (GC)}$ (MFR of the GCs) versus ${\cal R}_{\rm {E-I}}^{\rm {(syn)}}$ [E-I synapse ratio in Eq.~(\ref{eq:Rsyn})].
Here, open circles and crosses represent firing activities of GCs in the presence of only the external inputs from the EC and the HIPP cells [shown in Fig.~\ref{fig:EC-GC}(a)] and in the whole network (considering the additional MC effect) [shown in Fig.~\ref{fig:WN}(a)], respectively. Numbers denote the GC index $j$ in the $I=1$ GC cluster. In the case of considering only the external EC and HIPP inputs, there are 6 active GCs (represented by open circles).
Among them, only one $j=4$ GC (represented by the downarrow) with the highest ${\cal R}_{\rm {E-I}}^{\rm {(syn)}}$ survives and becomes the winner under the feedback inhibition from the BC; all the other 5 GCs with lower ${\cal R}_{\rm {E-I}}^{\rm {(syn)}}$ become silent. The spike train of the winner GC is shown in Fig.~\ref{fig:k1WTA}(a2). Among the 96 GC clusters with $k=1$ WTA competition, GCs with the highest ${\cal R}_{\rm {E-I}}^{\rm {(syn)}}$ become winners in the 89 GC clusters, as in the case of the $I=1$ GC cluster.

\begin{figure}
\includegraphics[width=0.8\columnwidth]{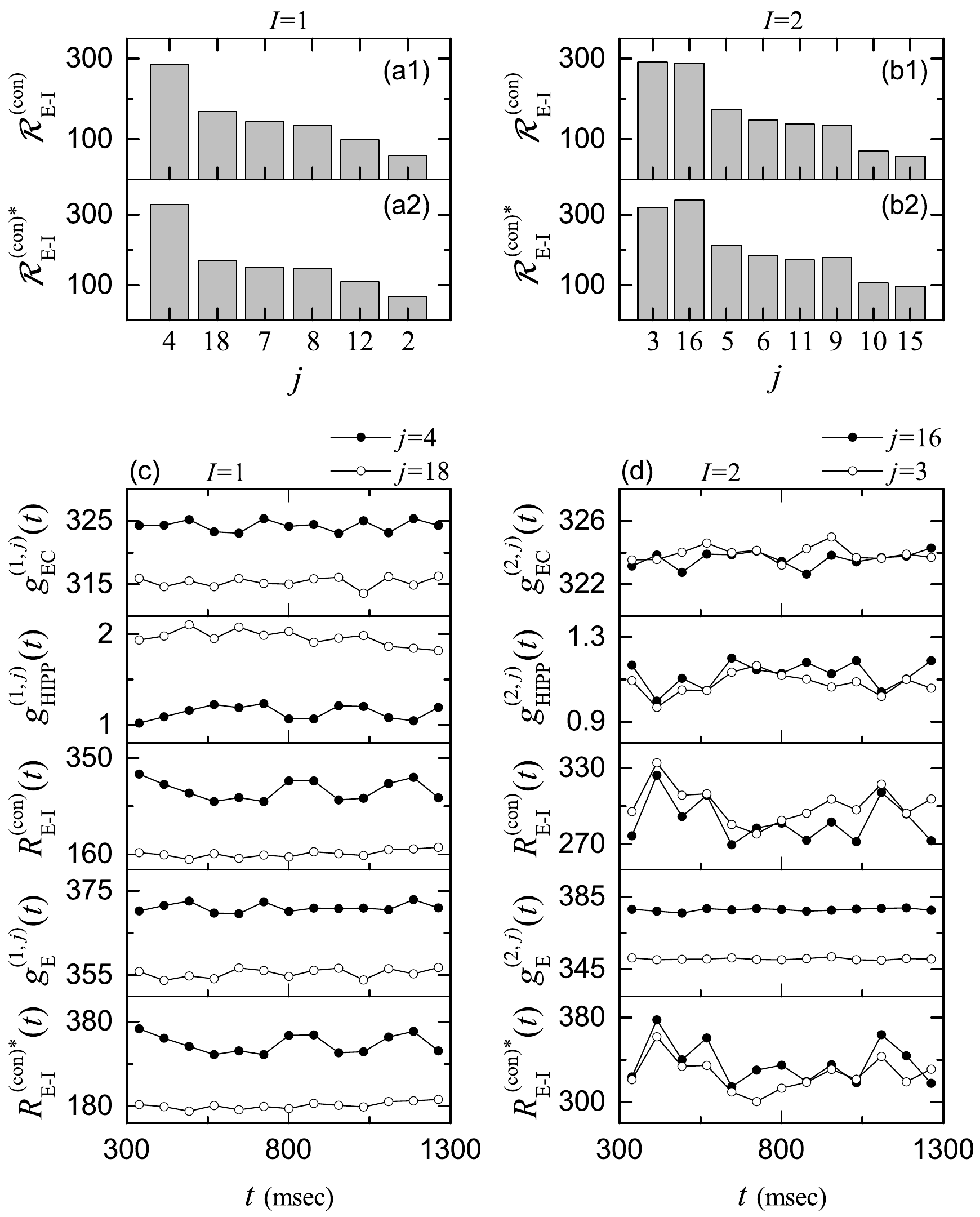}
\caption{Competition between the external excitatory and inhibitory inputs into the GCs in the case of the $k = 1$ WTA competition. Plots of (a1)
${\cal R}_{\rm E-I}^{\rm (con)}$  and (a2) ${ {\cal{R}}_{\rm E-I}^{\rm (con)}}^*$ of the 6 GCs [corresponding to the  active GCs in the presence of only the excitatory EC and the inhibitory HIPP inputs; see Fig.~\ref{fig:k1WTA}(a1)]  in the 1st ($I = 1$) GC cluster; $j$ ($x$-axis label) represents the GC index. Plots of (b1) ${\cal R}_{\rm E-I}^{\rm (con)}$  and (b2) ${ {\cal{R}}_{\rm E-I}^{\rm (con)}}^*$ of the 8 GCs [corresponding to the active GCs in the presence of only the excitatory EC and the inhibitory HIPP inputs; see Fig.~\ref{fig:k1WTA}(b1)] in the 2nd ($I = 2$) GC cluster.
(c) Time courses of $g_{\rm EC}^{(1,j)} (t)$, $g_{\rm HIPP}^{(1,j)}(t)$, $R_{\rm E-I}^{\rm (con)}(t)$, $g_{\rm E}^{(1,j)}(t)$, and
${R_{\rm E-I}^{\rm (con)}}^*(t)$ of the winner GC ($j=4$; solid circles) and the silent GC ($j=18$; open circles) with the 2nd highest
${ {\cal{R}}_{\rm E-I}^{\rm (con)}}^*$ in the 1st ($I=1$) GC cluster. (d) Time courses of $g_{\rm EC}^{(2,j)} (t)$, $g_{\rm HIPP}^{(2,j)}(t)$, $R_{\rm E-I}^{\rm (con)}(t)$, $g_{\rm E}^{(2,j)}(t)$, and ${R_{\rm E-I}^{\rm (con)}}^*(t)$ of the winner GC ($j=16$; solid circles) and the silent GC ($j=3$; open circles) with the 2nd highest ${ {\cal{R}}_{\rm E-I}^{\rm (con)}}^*$ in the 2nd ($I=2$) GC cluster.
}
\label{fig:DOk1}
\end{figure}

In the remaining 7 GC clusters ($I=$2, 34, 48, 61, 64, 86, and 89), GCs with the second highest ${\cal R}_{\rm {E-I}}^{\rm {(syn)}}$ become winners.
As an example, we consider the case of the $I=2$ GC cluster, which is shown in Fig.~\ref{fig:k1WTA}(b1). There are 8 active GCs (denoted by open circles along with the numbers) in the presence of only the external EC and HIPP inputs. In this case, the values of ${\cal R}_{\rm {E-I}}^{\rm {(syn)}}$ for the first ($j=3$) and the second ($j=16$) highest ones have small difference. We note again that ${\cal R}_{\rm {E-I}}^{\rm {(syn)}}$ represents only the ratio of the external excitatory input from the EC to the external inhibitory input from the HIPP cells without consideration of the effect of the MCs. When considering the effect of the MCs, the $j=16$ GC with the second highest ${\cal R}_{\rm {E-I}}^{\rm {(syn)}}$ has a larger ratio of the external excitatory input to the external inhibitory input than the $j=3$ GC with the first highest ${\cal R}_{\rm {E-I}}^{\rm {(syn)}}$, which will be shown in Fig.~\ref{fig:DOk1}. Thus, the $j=16$ GC becomes the winner in the $I=2$ GC cluster. Its spike train is shown in Fig.~\ref{fig:k1WTA}(b2).

We now investigate the dynamical origin of the $k=1$ WTA competition.
WTA competition occurs via competition between the firing activity of the GCs and the feedback inhibition from the BC.
Then, only strongly active GCs may survive under the feedback inhibition.
In this case, the firing activity of the GCs is determined through competition between the external excitatory inputs from the EC
cells and the MCs and the external inhibitory input from the HIPP cells.
When the magnitude of the excitatory synaptic currents is sufficiently larger than that of the inhibitory synaptic current,
the firing activity of the GCs becomes strong.

As in Eq.~(\ref{eq:ISyn2}), synaptic current is given by the product of synaptic conductance $g$ and potential difference. In this case, synaptic conductance determines the time course of the synaptic current. Hence, it is sufficient to consider the time-course of synaptic conductance. The
synaptic conductance $g$ is given by the product of synaptic strength per synapse ($K$), the number of synapses ($M_{\rm syn})$, and the
fraction $s$ of open (post-synaptic) ion channels [see Eq.~(\ref{eq:ISyn3})].

As in Eq.~(\ref{eq:ISyn4}), time course of $s(t)$ is given by the summation for double-exponential functions over pre-synaptic spikes.
Here, we make an approximation of the fraction $s(t)$ of open ion channels (i.e., contributions of summed effects of pre-synaptic spikes) by the bin-averaged spike rate $f_X^{(I,j)}(t)$ of pre-synaptic neurons in the $X$-population innervating the $(I,j)$ GC (i.e., $j$th GC in the $I$th GC cluster), as in our previous case of cerebellar Pavlovian eyeblink conditioning \cite{Kim2}.
Then, the excitatory conductance $g_{\rm {EC}}^{(I,j)}$ for the synaptic current $I_{\rm {syn}}^{(I,j)}$ from the pre-synaptic EC cells into the
$(I,j)$ GC (i.e., the $j$th GC in the $I$th GC cluster) is given by:
\begin{eqnarray}
g_{\rm {EC}}^{(I,j)}(t) &=& g_{\rm {EC,AMPA}}^{(I,j)}(t) + g_{\rm {EC,NMDA}}^{(I,j)}(t) \nonumber \\
                     &\simeq& (K_{\rm {AMPA}} ^{\rm {(GC,EC)}} + K_{\rm {NMDA}}^{\rm {(GC,EC)}}) \\
                     & & \times M_{\rm {syn}}^{\rm {(GC,EC)}}
                     f_{\rm {EC}}^{(I,j)}(t). \nonumber
\label{eq:ECCon1}
\end{eqnarray}
Here, the values of $K_{\rm {AMPA}} ^{\rm {(GC,EC)}}$ and  $K_{\rm {NMDA}}^{\rm {(GC,EC)}}$ are given in Table \ref{tab:Synparm1} and the bin-averaged
spike rate $f_{\rm EC}^{(I,j)}(t)$ of pre-synaptic EC cells in the $i$th bin is given by:
\begin{equation}
f_{\rm EC}^{(I,j)}(t) = {\frac {N_i^{(s)}(t)} {N_{pre}^{(I,j,{\rm EC})}~ \Delta t}},
\label{eq:SR}
\end{equation}
where $N_i^{(s)}(t)$ is the number of spikes of the pre-synaptic EC cells in the $i$th bin, $N_{pre}^{(I,j,{\rm EC})}$ is the number of the pre-synaptic EC cells
innervating the $(I,j)$ GC neuron, and $\Delta t$ (= 77 msec) is the bin size.
Thus, we obtain the excitatory conductance $g_{\rm {EC}}^{(I,j)}$:
\begin{equation}
g_{\rm {EC}}^{(I,j)}(t) \simeq 1.04~ M_{\rm {syn}}^{\rm {(GC,EC)}} f_{\rm {EC}}^{(I,j)}(t).
\label{eq:ECCon2}
\end{equation}
Similarly, we also get the inhibitory conductance $g_{\rm {HIPP}}^{(I,j)}$ for the synaptic current $I_{\rm {syn}}^{(I,j)}$ from the pre-synaptic HIPP cells into the
$(I,j)$ GC:
\begin{equation}
 g_{\rm {HIPP}}^{(I,j)}(t) = g_{\rm {HIPP,GABA}}^{(I,j)}(t) \simeq 0.12~ M_{\rm {syn}}^{\rm {(GC,HIPP)}} f_{\rm {HIPP}}^{(I,j)}(t).
\label{eq:HIPPCon}
\end{equation}

In Figs.~\ref{fig:E-ICom} and \ref{fig:k1WTA}, we consider only $M_{\rm {syn}}$ (number of synapses) as a ``simplified'' version of the synaptic input.
In contrast, we now consider the ``full'' version of the synaptic conductance $g$ by taking into consideration additional synaptic strength $K$ and
bin-averaged spike rate $f$ together with $M_{\rm {syn}}$. Then, the ratio of the external excitatory to inhibitory conductance is given by:
\begin{equation}
  R_{\rm {E-I}}^{\rm {(con)}}(t) = {\frac {g_{\rm {EC}}^{(I,j)}(t)} {g_{\rm {HIPP}}^{(I,j)}(t)}}.
\label{eq:RCon1}
\end{equation}
In this case, we introduce the E-I conductance ratio ${\cal R}_{\rm E-I}^{\rm (con)}$, defined by the time average of $R_{\rm {E-I}}^{\rm {(con)}}(t)$:
\begin{equation}
 {\cal R}_{\rm E-I}^{(con)} = \overline{ R_{\rm {E-I}}^{\rm {(con)}}(t) } = \overline{  {\frac {g_{\rm {EC}}^{(I,j)}(t)} {g_{\rm {HIPP}}^{(I,j)}(t)}} },
\label{eq:RCon}
\end{equation}
where the overline denotes time average.

The E-I conductance ratio ${\cal R}_{\rm E-I}^{\rm (con)},$ representing the time-averaged ratio of the excitatory EC to the inhibitory HIPP conductances,
corresponds to a refined version in comparison to the E-I synapse ratio ${\cal R}_{\rm E-I}^{\rm (syn)}$ in Eq.~(\ref{eq:Rsyn}).
Figures \ref{fig:DOk1}(a1) and \ref{fig:DOk1}(b1) show the histograms of the E-I conductance ratio ${\cal R}_{\rm E-I}^{(con)}$ versus the active GCs
(in the presence of only the excitatory EC and the inhibitory HIPP inputs) in the $I=1$ and 2 GC clusters, respectively; such active GCs are denoted by open circles
with numbers in Figs.~\ref{fig:k1WTA}(a1) and \ref{fig:k1WTA}(b1).
We note that the order of the magnitude of ${\cal R}_{\rm E-I}^{\rm (con)}$ is the same as that for ${\cal R}_{\rm E-I}^{\rm (syn)}$. Particularly, in the case of the $I=2$ GC cluster, the 1st highest ${\cal R}_{\rm E-I}^{\rm (con)}$ for the $j=3$ active GC is just a little larger than the 2nd highest
${\cal R}_{\rm E-I}^{\rm (con)}$ for the $j=16$ active GC. However, the order becomes reversed when we consider the effect of the excitatory input from the MCs.

We now include the external excitatory MC input whose conductance is given by:
\begin{eqnarray}
 g_{\rm {MC}}^{(I,j)}(t) &=& g_{\rm {MC,AMPA}}^{(I,j)}(t) + g_{\rm {MC,NMDA}}^{(I,j)}(t) \nonumber \\
 &\simeq& 0.06~M_{\rm {syn}}^{\rm {(GC,MC)}} f_{\rm {MC}}^{(I,j)}(t).
\label{eq:MCCon}
\end{eqnarray}
Then, we get the total excitatory input $g_E^{(I,j)}(t)$ via adding $g_{\rm {EC}}^{(I,j)}(t)$ and  $g_{\rm {MC}}^{(I,j)}(t)$;
\begin{equation}
g_E^{(I,j)}(t) = g_{\rm {EC}}^{(I,j)}(t) + g_{\rm {MC}}^{(I,j)}(t).
\label{TEcon}
\end{equation}
In this case, the ratio of the total external excitatory to inhibitory conductance is given by:
\begin{equation}
  {R_{\rm {E-I}}^{\rm {(con)}}}^*(t) = {\frac {g_E^{(I,j)}(t)} {g_{\rm {HIPP}}^{(I,j)}(t)}}.
\label{eq:RCon2}
\end{equation}
Then, in the whole network (including the MCs), we introduce the E-I conductance ratio ${ {\cal{R}}_{\rm E-I}^{\rm (con)}}^*$, given by the time average of ${R_{\rm {E-I}}^{\rm {(con)}}}^*(t)$:
\begin{equation}
  { {\cal{R}}_{\rm E-I}^{\rm (con)}}^* = \overline{ {R_{\rm {E-I}}^{\rm {(con)}}}^*(t) }
  = \overline{  {\frac {g_{\rm {E}}^{(I,j)}(t)} {g_{\rm {HIPP}}^{(I,j)}(t)}} }.
\label{eq:RCon*}
\end{equation}
The E-I conductance ratio ${ {\cal{R}}_{\rm E-I}^{\rm (con)}}^*$ (considering the MC effect) represents the ratio of the external excitatory to the inhibitory inputs
better than the E-I conductance ratio ${\cal R}_{\rm E-I}^{\rm (con)}$ (in the presence of only the excitatory EC and the inhibitory HIPP inputs).
Hence, ${ {\cal{R}}_{\rm E-I}^{\rm (con)}}^*$ becomes a more refined version
in comparison to ${\cal R}_{\rm E-I}^{\rm (con)}$ (which does not consider the effect of the MCs).

Figures \ref{fig:DOk1}(a2) and \ref{fig:DOk1}(b2) show the histograms of the (refined) E-I conductance ratio ${ {\cal{R}}_{\rm E-I}^{\rm (con)}}^*$ (considering the MC effect) versus the active GCs in the $I=1$ and 2 GC clusters, respectively. In the case of the $I=1$ case, the order of the magnitude of
${ {\cal{R}}_{\rm E-I}^{\rm (con)}}^*$ is the same as that of ${\cal{R}}_{\rm E-I}^{\rm (con)}$, and hence the $j=4$ GC continues to become the winner.
On the other hand, in the case of $I=2$ GC cluster, the $j=16$ GC (with the second highest $ {\cal{R}}_{\rm E-I}^{\rm (con)}$)
has a higher ${ {\cal{R}}_{\rm E-I}^{\rm (con)}}^*$ than the $j=3$ GC (with the first highest ${\cal{R}}_{\rm E-I}^{\rm (con)}$)).
Thus, the $j=16$ GC with the highest ${ {\cal{R}}_{\rm E-I}^{\rm (con)}}^*$ becomes the winner.

\begin{figure}
\includegraphics[width=0.9\columnwidth]{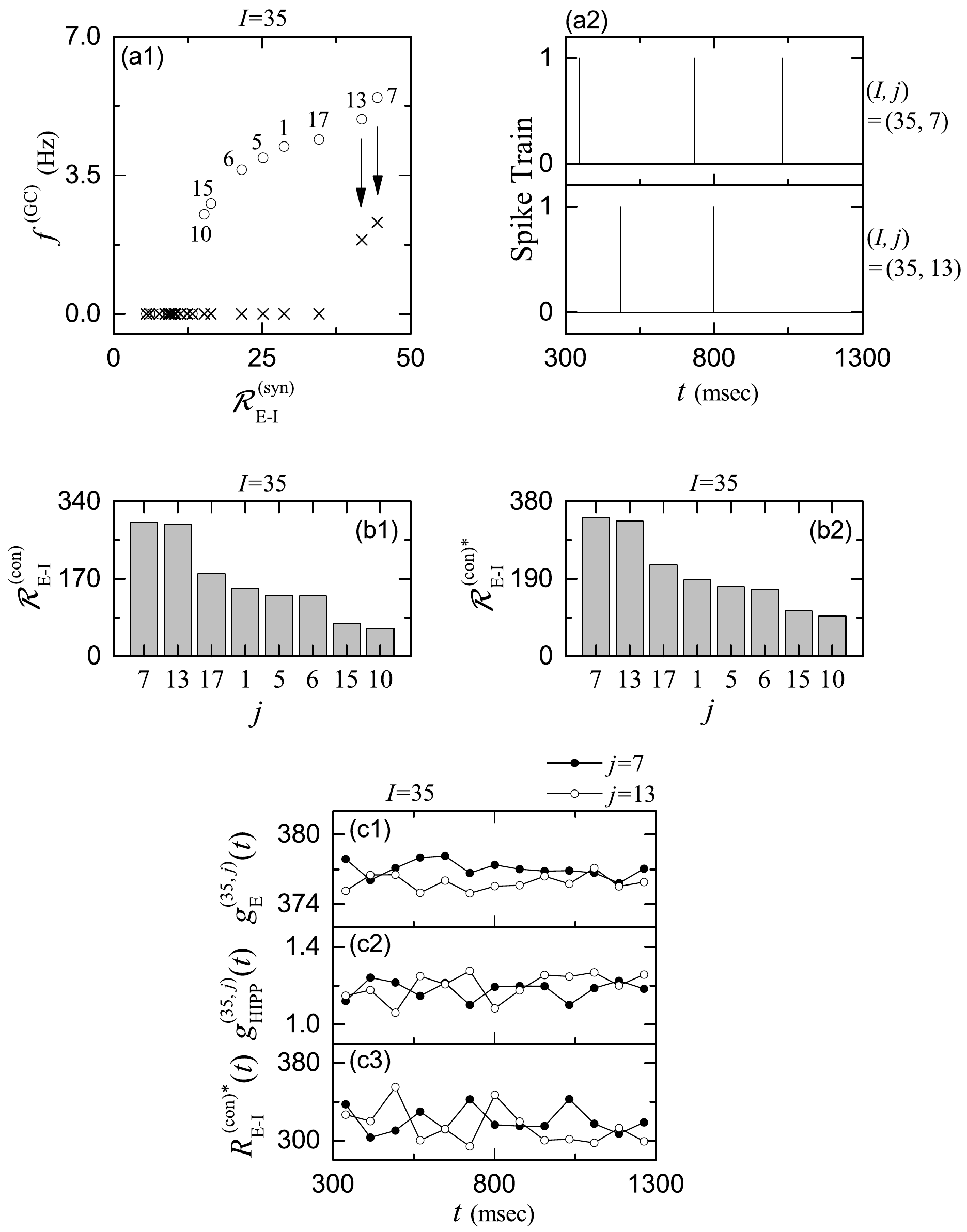}
\caption{Competition between the external excitatory and inhibitory inputs into the GCs in the case of the $k = 2$ WTA competition. (a1) Plot of $f^{\rm (GC)}$
(MFR of GCs) versus the E-I synapse ratio ${\cal R}_{\rm E-I}^{\rm (syn)}$ in the 35th ($I = 35$) GC cluster. Open circles and crosses represent firing activities of GCs in the presence of only the external inputs from the EC and the HIPP cells [shown in Fig.~\ref{fig:EC-GC}(a)] and in the whole network [shown in Fig.~\ref{fig:WN}(a)], respectively; numbers represent the GC index. (a2) Spike trains of the two winner GCs ($j = 7$ and 13) in the 35th ($I = 35$) GC cluster. Plots of (b1) ${\cal R}_{\rm E-I}^{\rm (con)}$  and (a2) ${ {\cal{R}}_{\rm E-I}^{\rm (con)}}^*$ of the 8 GCs [corresponding to the active GCs in the presence of only the excitatory EC input and the inhibitory HIPP input; see Fig.~\ref{fig:k2WTA}(a1)] in the 35th ($I = 35$) GC cluster; $j$ ($x$-axis label) denotes the GC index. Time courses of (c1) $g_{\rm E}^{(35,j)}(t)$, (c2) $g_{\rm HIPP}^{(35,j)}(t)$, and (c3) ${R_{\rm E-I}^{\rm (con)}}^*(t)$ of the two winner GCs; $j=7$ (solid circles) with the highest ${ {\cal{R}}_{\rm E-I}^{\rm (con)}}^*$ and $j=13$ (open circles) with the 2nd highest ${ {\cal{R}}_{\rm E-I}^{\rm (con)}}^*$ in the 35th ($I=35$) GC cluster.
}
\label{fig:k2WTA}
\end{figure}

Figure \ref{fig:DOk1}(c) shows the time-evolutions of $g_{\rm {EC}}^{(I,j)}(t)$, $g_{\rm {HIPP}}^{(I,j)}(t)$, $R_{\rm {E-I}}^{\rm {(con)}}(t),$
$g_{\rm {E}}^{(I,j)}(t)$, and ${R_{\rm {E-I}}^{\rm {(con)}}}^*(t)$ for the winner GC of $j=4$ and the silent GC of $j=18$ with the second highest
${ {\cal{R}}_{\rm E-I}^{\rm (con)}}^*$ in the $I=1$ GC cluster. The total excitatory conductance $g_{\rm {E}}^{(I,j)}(t)$
for the $j=4$ GC is larger than that for the $j=18$ GC, while the inhibitory conductance $g_{\rm {HIPP}}^{(I,j)}(t)$
for for the $j=4$ GC is smaller than that for the $j=18$ GC. Hence, the ratio ${R_{\rm {E-I}}^{\rm {(con)}}}^*(t)$ for the winner ($j=4$ GC) becomes
greater than that for the $j=18$ GC.

We next consider the time-evolutions of $g_{\rm {EC}}^{(I,j)}(t)$, $g_{\rm {HIPP}}^{(I,j)}(t)$, $R_{\rm {E-I}}^{\rm {(con)}}(t),$
$g_{\rm {E}}^{(I,j)}(t)$, and ${R_{\rm {E-I}}^{\rm {(con)}}}^*(t)$ for the winner GC of $j=16$ and the silent GC of $j=3$
with the second highest ${ {\cal{R}}_{\rm E-I}^{\rm (con)}}^*$   in the $I=2$ GC cluster in Fig.~\ref{fig:DOk1}(d).
The $j=3$ GC tends to have a larger excitatory EC conductance $g_{\rm {EC}}^{(I,j)}(t)$ than the $j=16$ winner GC, and it also tends to have
a smaller inhibitory HIPP conductance $g_{\rm {HIPP}}^{(I,j)}(t)$ than the $j=16$ winner. Thus, their ratio $R_{\rm {E-I}}^{\rm {(con)}}(t)$
becomes larger for the $j=3$ GC. However, when including the excitatory MC conductance,  the $j=16$ GC has much higher total excitatory conductance
$g_{\rm {E}}^{(I,j)}(t)$ than the $j=3$ GC. Thus, the $j=16$ GC becomes the winner in the $I=2$ GC cluster.

In addition to the $k=1$ WTA competition, we also consider the $k=2$ WTA competition.
Figure \ref{fig:k2WTA}(a1) shows the WTA competition in the $I=35$ GC cluster with the two winners ($j=7$ and 13).
A plot of $f^{\rm {(GC)}}$ (MFRs of the GCs) versus the E-I synapse ratio ${\cal R}_{\rm {E-I}}^{\rm (syn)}$ is shown. There are 8 active GCs (denoted by open circles with the numbers) in the presence of only the excitatory EC and the inhibitory HIPP inputs.
Among them, the $j=7$ and 13 GCs become the winners in the whole DG network. Their spike trains are shown in Fig.~\ref{fig:k2WTA}(a2).
These winners exhibit random intermittent spikings; their spikings also appear in different bins.

\begin{figure}
\includegraphics[width=0.9\columnwidth]{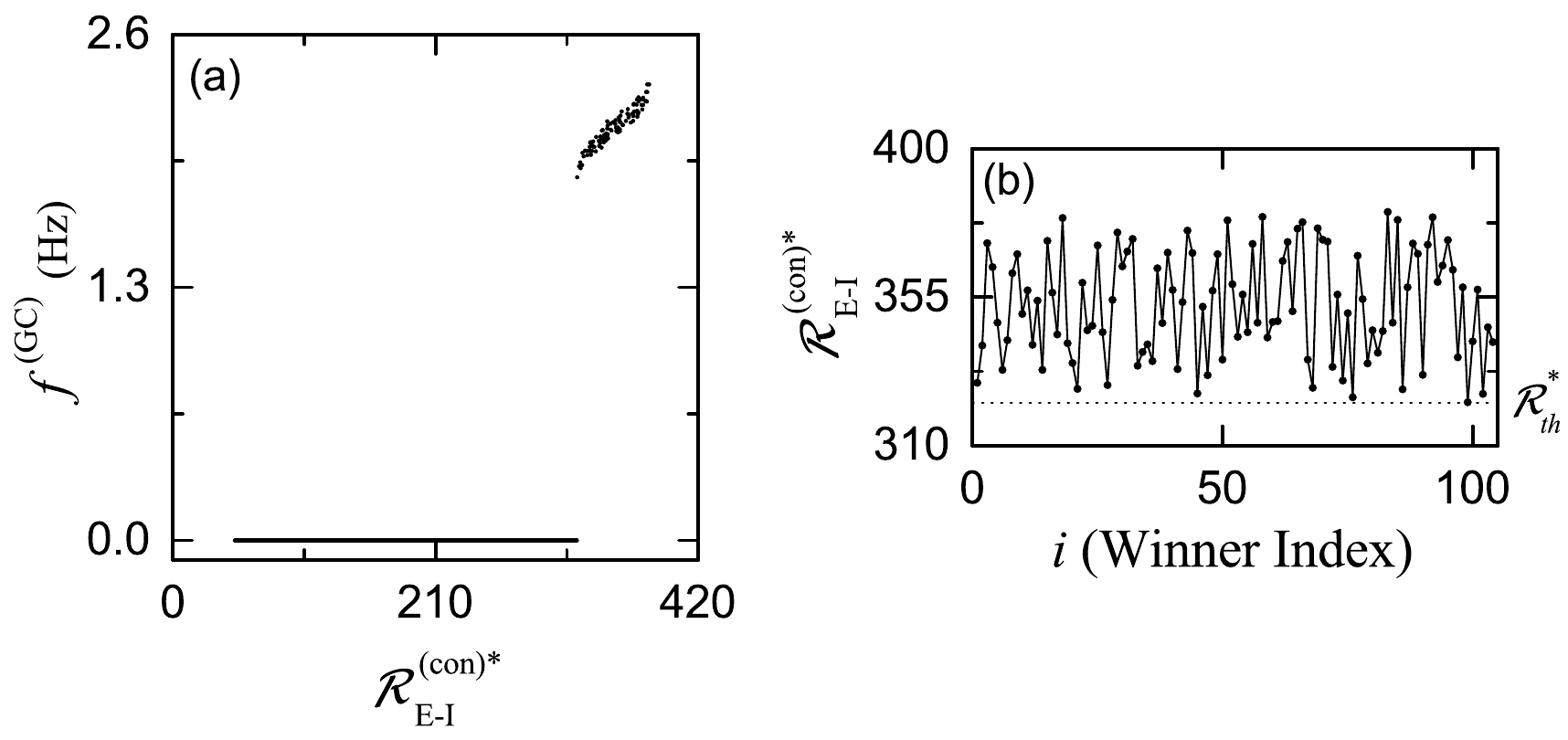}
\caption{Determination of the threshold ${\cal R}_{th}^*$ for the winner. (a) Plot of $f^{\rm (GC)}$ (MFR of GCs) versus
${ {\cal{R}}_{\rm E-I}^{\rm (con)}}^*$ (E-I conductance ratio which considers the effect of the MCs). (b) Plot of ${ {\cal{R}}_{\rm E-I}^{\rm (con)}}^*$  versus $i$ (index of the winner GCs). Horizontal dotted line denotes the threshold ${\cal R}_{th}^*~(=323.21)$ for the winner.
}
\label{fig:Wth}
\end{figure}

To investigate the dynamical origin of the $k=2$ WTA competition, we consider the E-I conductance ratios,
${\cal{R}}_{\rm E-I}^{\rm (con)}$ (without considering the MC effect) in Eq.~(\ref{eq:RCon}) and ${ {\cal{R}}_{\rm E-I}^{\rm (con)}}^*$
(considering the MC effect) in Eq.~(\ref{eq:RCon*}), representing the competition between the external excitatory and the inhibitory conductance.
Their histograms versus the 8 active GCs are shown in Figs.~\ref{fig:k2WTA}(b1) and \ref{fig:k2WTA}(b2), respectively.
The order of magnitude of the ratio is the same in both cases of ${\cal{R}}_{\rm E-I}^{\rm (con)}$ and ${ {\cal{R}}_{\rm E-I}^{\rm (con)}}^*$.
We note that the first and the second highest cases for $j=7$ and 13 are nearly the same, and they are distinctly larger than the 3rd highest one ($j=17$). For example, the values of ${ {\cal{R}}_{\rm E-I}^{\rm (con)}}^*$  for $j=7$ and 13 are 340.65 and 335,58, respectively. Since these values are larger than the threshold ${\cal R}_{th}^*~(=323)$ in Fig.~\ref{fig:Wth}, both the $j=7$ and 13 GCs become the winners.

Figures \ref{fig:k2WTA}(c1)-\ref{fig:k2WTA}(c3) show the time-evolutions of $g_{\rm {E}}^{(I,j)}(t)$, $g_{\rm {HIPP}}^{(I,j)}(t)$, and
${R_{\rm {E-I}}^{\rm {(con)}}}^*(t)$ for the $j=7$ and 13 GCs, respectively. The $j=7$ GC tends to have a little larger excitatory conductance $g_{\rm {E}}^{(I,j)}(t)$. On the other hand, they have nearly the same time-average of the inhibitory conductance $g_{\rm {HIPP}}^{(I,j)}(t)$ which shows alternative up-and-down evolutions. Thus, the ratios ${R_{\rm {E-I}}^{\rm {(con)}}}^*(t)$ for the $j=7$ and 13 GCs also show alternative up-and-down evolutions due to the inhibitory conductance, and they have only a little different time-averages. In Fig.~\ref{fig:k2WTA}(c3), the $j=7$ (13) GC enter the up-stage three times (twice).
In the up stage, the corresponding GC fires a spiking. Hence, in Fig.~\ref{fig:k2WTA}(a2), 3 (2) spikes appear in the spike train of the $j=7$ (13) GC.

Figure \ref{fig:Wth}(a) shows the plot of $f^{\rm {(GC)}}$ (MFRs of all the GCs) versus ${ {\cal{R}}_{\rm E-I}^{\rm (con)}}^*$.
We determine the threshold ${\cal R}_{th}^*$ for the E-I conductance ratio ${ {\cal{R}}_{\rm E-I}^{\rm (con)}}^*$ (considering the MC effect).
We note that active winner GCs with $f^{\rm {(GC)}}>0$ appear when ${ {\cal{R}}_{\rm E-I}^{\rm (con)}}^*$
passes a threshold ${\cal R}_{th}^*~(=323.21)$. Figure \ref{fig:Wth}(b) shows a plot of ${ {\cal{R}}_{\rm E-I}^{\rm (con)}}^*$ versus 104 winner GCs. The range of
${ {\cal{R}}_{\rm E-I}^{\rm (con)}}^*$ is [${ {\cal{R}}_{\rm E-I}^{\rm (con)}}^* {\rm (min)}$, ${ {\cal{R}}_{\rm E-I}^{\rm (con)}}^* {\rm (max)}$];
${ {\cal{R}}_{\rm E-I}^{\rm (con)}}^* {\rm (min)} = 323.21$ and ${ {\cal{R}}_{\rm E-I}^{\rm (con)}}^* {\rm (max)}= 380.82$.
Then, we get the winner threshold percentage $W_{th}\%~(=15.1 \%)$:
\begin{equation}
W_{th}\% = \frac { [ { {\cal{R}}_{\rm E-I}^{\rm (con)}}^* {\rm (max)} - { {\cal{R}}_{\rm E-I}^{\rm (con)}}^* {\rm (min)} ] }
           { { {\cal{R}}_{\rm E-I}^{\rm (con)}}^* {\rm (max)} } \times 100
\label{eq:Wth}
\end{equation}
Thus, active winner GCs have their ${ {\cal{R}}_{\rm E-I}^{\rm (con)}}^*$  within $W_{th}\%$ of the maximum
${ {\cal{R}}_{\rm E-I}^{\rm (con)}}^* {\rm (max)}$ of the GC with the strongest activity
[i.e., GCs become active winners when their ${ {\cal{R}}_{\rm E-I}^{\rm (con)}}^*$ lies within $W_{th}\%$ of the maximum
${ {\cal{R}}_{\rm E-I}^{\rm (con)}}^* {\rm (max)}$].

\subsection{Effect of The Hilar MCs on The WTA Competition}
\label{subsec:MC}
Finally, we are concerned about the effect of the hilar MCs on the WTA competition. As shown in Fig.~\ref{fig:WN}(b), the MCs facilitate firing activity of the GC-BC loop by exciting both the GCs and the BCs. However, MC loss may occur during epileptogenesis \cite{BN1,BN2}, which may be a cause of impaired pattern separation leading to memory interference. We investigate the role of the MCs for the WTA competition through their ablation.

We consider the case of complete MC loss. Figures \ref{fig:MC1}(a1)-\ref{fig:MC1}(e) show the WTA competition in the DG network without the MCs.
In this case, the BC activity becomes weakened, which leads to decrease in the feedback inhibition to the GCs.
Then, the GC activity becomes strengthened. Thus, more winner GCs appear, as shown in Fig.~\ref{fig:MC1}(a1) showing the plot of
$k$ (number of winner GCs) versus $I$ (GC cluster index). The range of $k$ is [2, 8], which is much widened in comparison to the original case
[see Fig.~\ref{fig:WN}(b)] in the whole network with the MCs. The total number of active GCs is 502, and hence the activation degree
$D_a$ is $25.1\%$, which is much larger than $D_a=5.2\%$ in the whole network with the MCs. Figure \ref{fig:MC1}(a2) also shows the plot of the number of the GC clusters versus $k$. The $k=5$ corresponds to the most probable case where the number of the corresponding GC clusters is 27, and the mean value of $k$ is also
5.06.

As an example, we consider the $I=1$ GC cluster. Figure \ref{fig:MC1}(b) shows the plot of $f^{\rm (GC)}$ (MFR of the GCs) versus the E-I synapse ratio
${\cal R}_{\rm E-I}^{\rm (syn)}$. Only in the presence of the excitatory EC input and the inhibitory HIPP input,
there are 6 active GCs, denoted by open circles with the numbers, among which three winners of $j=$4, 18, and 7 survive in response to the feedback inhibition from the BC, in contrast to the case of the whole network (including the MCs) in Fig.~\ref{fig:k1WTA}(a1) where only the $j=4$ GC is the winner. Spike trains of these active GCs are shown in Figs.~\ref{fig:MC1}(c1)-\ref{fig:MC1}(c3), respectively. They exhibit stochastic intermittent spikings; their spikings appear in different bins.

Figure \ref{fig:MC1}(d) shows the plot of ${ {\cal{R}}_{\rm E-I}^{\rm (con)}}^*$ in Eq.~(\ref{eq:RCon*}) (representing the time-averaged ratio of the external excitatory to inhibitory conductance) of the 6 active GC cells [$j=4$, 18, 7, 8, 12, and 2 in Fig.~\ref{fig:MC1}(b)] in the presence of only the excitatory EC and the inhibitory HIPP inputs (without MCs). Here, the values of ${ {\cal{R}}_{\rm E-I}^{\rm (con)}}^*$ for the first three GCs ($j=4$, 18, and 7) are 285.60, 170.10, and 145.63, respectively; in the case of the 4th $j=8$ GC, ${ {\cal{R}}_{\rm E-I}^{\rm (con)}}^* = 132.$

Figure \ref{fig:MC1}(e) shows the plot of $f^{(GC)}$ (MFR of all the GCs) versus ${ {\cal{R}}_{\rm E-I}^{\rm (con)}}^*$.
When ${ {\cal{R}}_{\rm E-I}^{\rm (con)}}^*$ passes a threshold ${\cal R}_{th}^*~(=134.57)$, active GCs with $f^{\rm (GC)} >0$ appear.
The range of ${ {\cal{R}}_{\rm E-I}^{\rm (con)}}^*$ for the active GCs is [134.57, 298.96].
Hence, in the case of complete loss of MCs, the winner threshold percentage $W_{th}\%$ in Eq.~(\ref{eq:Wth}) becomes $55.0\%$, which is much larger than
$W_{th}^*~(=15.1\%)$ in the whole network with MCs (see Fig.~\ref{fig:Wth}). Thus, active winner GCs have their
${ {\cal{R}}_{\rm E-I}^{\rm (con)}}^*$ within $W_{th}\%~(=55\%)$ of the maximum ${ {\cal{R}}_{\rm E-I}^{\rm (con)}}^* {\rm {(max)}} $ of the GC with the strongest activity. Consequently, in Fig.~\ref{fig:MC1}(d), the first three GCs of $j=4$, 18, and 7 become the winners; the 4th GC of $j=8$ becomes silent. In comparison to the case of the whole network with the MCs, two more winners of $j=18$ and 7 appear (i.e., the WTA competition becomes weakened) because the firing activity of the GCs becomes strengthened in the case without the MCs.

We also decrease $N_{\rm MC}$ from 80 (in the original whole network) to 0 (complete loss). In this case, the fraction of MCs ($F_{\rm MC}$) is given by
$N_{\rm MC} / 80$. Figure \ref{fig:MC1}(f1) shows the plot of the activation degree $D_a$ of the GCs versus $F_{\rm MC}$.
As $F_{\rm MC}$ is decreased from 1 to 0, the activity of the BCs becomes weakened, which results in decrease in the feedback inhibition to the
GCs. Then, the activity of the GCs becomes strengthened (i.e., $D_a$ increases monotonically from $5.2\%$ to $25.1\%$).
In this case, the winner threshold percentage $W_{th}\%$ also increases from $15.1\%$ to $55.0\%$ with decreasing $F_{\rm MC}$ from 1 to 0,
as shown in Fig.~\ref{fig:MC1}(f2). Due to the increased $W_{th}\%$ more active MCs appear. Thus, WTA competition becomes more and more weakened with decreasing $F_{\rm MC}$. The WTA competition becomes the strongest in the case of $F_{\rm MC}=1$ where the firing activity of the GCs is the most sparse.
In this way, the role of the MCs is essential for the WTA competition. Finally, Fig.~\ref{fig:MC1}(f3) shows the plot of $W_{th}\%$ versus $D_a$. There exists a positive correlation (with the Pearson's correlation coefficient $r=0.9884$) between $W_{th}\%$ and $D_a$; a fitted dashed line is given. The larger the activation degree $D_a$ of the GCs is, the higher the winner threshold percentage $W_{th}\%$ becomes.

\section{Summary and Discussion}
\label{sec:SUM}
We considered a biological network of the hippocampal DG, and investigated the dynamical origin of the WTA competition leading to sparse activation of the GCs.
Such sparsity has been known to improve the pattern separation (pre-processed in the DG) to facilitate the pattern storage and retrieval in the CA3.
In each GC cluster, a dynamical GC-BC loop is formed; all the excitatory GCs are synaptically coupled with the single inhibitory BC.
Active GC winners are selected via competition between the firing activity of the GCs and the feedback inhibition of the BC.
Only strongly active GCs may survive in response to the feedback inhibition from the BC.

The EC and the hilar MCs are the external input sources to the GCs. Thus, there are three types of external inputs into the GCs; the direct excitatory EC input,
the indirect inhibitory EC input, mediated by the HIPP cells, and the excitatory input from the MCs.
Then, the firing activities of the GCs are determined through competition between the external E and I inputs to the GCs; two excitatory inputs from the EC via
the PPs and from the MCs and one inhibitory input from the HIPP cells. It has been shown that the degree of the external E-I input competition may be well represented by the E-I conductance ratio ${ {\cal{R}}_{\rm E-I}^{\rm (con)}}^*$ (given by the time average of the external E to I conductances).
GCs were found to become active when their ${ {\cal{R}}_{\rm E-I}^{\rm (con)}}^*$ is greater than the threshold ${\cal R}_{th}^*$, and their MFRs were also found to be strongly correlated with ${ {\cal{R}}_{\rm E-I}^{\rm (con)}}^*$. In this way, we have characterized the degree of the firing activity of the GCs in terms of ${ {\cal{R}}_{\rm E-I}^{\rm (con)}}^*$.

The WTA competition occurs through interaction of the firing activity of the GCs with the feedback inhibition of the BC.
GCs with larger ${ {\cal{R}}_{\rm E-I}^{\rm (con)}}^*$ than the threshold ${\cal R}_{th}^*$ have been found to survive under the feedback inhibition, and they became winners. On the other hand all the other GCs with smaller ${ {\cal{R}}_{\rm E-I}^{\rm (con)}}^*$ became silent in response to the feedback inhibition.
Among the 100 GC clusters, 96 clusters (corresponding to 96 $\%$) have been found to have only one winner ($k=1)$; the other 4 clusters had $k=2$ winners.
Thus, only 104 active GCs became the winners among the 2000 GCs , which corresponded to the activation degree $D_a=5.2$$\%$ (i.e., sparse activation).

We have also studied the WTA competition by varying $F_{\rm MC}$ (fraction of MCs). It was thus found that, as  $F_{\rm MC}$ is decreased, the activation degree $D_a$ of the GCs increases, the winner threshold percentage $W_{th}$$\%$  is also increased, and both $D_a$ and $W_{th}$$\%$ are strongly correlated.
We note that GCs become winners if their ${ {\cal{R}}_{\rm E-I}^{\rm (con)}}^*$ lies within $W_{th}$$\%$ of the maximum
$ {\cal{R}}_{\rm E-I,max}^{{\rm (con)}^*}$ of the GC with the strongest activity. Due to increased $W_{th}$$\%$, the number $k$ of the winner GCs
was found to increase. Thus, with decreasing $F_{\rm MC},$ the WTA competition became weakened. In this way, the hilar MCs play an important role to enhance the WTA competition in the GC-BC loop by providing excitation to both the GCs and the BC.

\begin{table}
\caption{In the LIF spiking neuron models, parameter values for the capacitance $C_X,$ the leaky current $I_L^{(X)},$ and the AHP current $I_{AHP}^{(X)}$
of the granule cell (GC) and the basket cell (BC) in the granular layer and the mossy cell (MC) and the hilar perforant path-associated (HIPP) cell in the hilus.
}
\label{tab:Singleparm}
\begin{tabular}{|c|c|c|c|c|c|}
\hline
\multicolumn{2}{|c|}{\multirow{2}{*}{$X$-population}} & \multicolumn{2}{c|}{Granular Layer} & \multicolumn{2}{c|}{Hilus} \\ \cline{3-6}
\multicolumn{2}{|c|}{} & \hspace*{0.5cm} GC \hspace*{0.5cm} & \hspace*{0.5cm} BC \hspace*{0.5cm} & \hspace{0.5cm} MC \hspace*{0.5cm}& HIPP cell \\
\hline
\multicolumn{2}{|c|}{$C_X$} & 106.2 & 232.6 & 206.0 & 94.3 \\ \hline
\multirow{2}{*}{$I_L^{(X)}$} & $g_L^{(X)}$ & 3.4 & 23.2 & 5.0 & 2.7 \\ \cline{2-6}
& $V_L^{(X)}$ & -75.0 & -62.0 & -62.0 & -65.0 \\ \cline{2-6}
\hline
\multirow{4}{*}{$I_{AHP}^{(X)}$} & $\bar{g}_{AHP}^{(X)}$ & 10.4 & 76.9 & 78.0 & 52.0 \\ \cline{2-6}
& $\tau_{AHP}^{(X)}$ & 20.0 & 2.0 & 10.0 & 5.0 \\ \cline{2-6}
& $V_{AHP}^{(X)}$ & -80.0 & -75.0 & -80.0 & -75.0 \\ \cline{2-6}
& $v_{th}^{(X)}$ & -53.4 & -52.5 & -32.0 & -9.4 \\ \cline{2-6}
\hline
\end{tabular}
\end{table}

Finally, we discuss limitations of our present work and future works.
In the present work, we considered only the case of ablating the MCs for studying their role. As a future work, it would also be interesting to investigate the WTA competition by changing the synaptic strength $K_R^{\rm (BC,MC)}$ $(R=\rm{NMDA~and~AMPA})$ of the synapses between the pre-synaptic MCs and the post-synaptic BC. The effect of decreasing $K_R^{\rm (BC,MC)}$ would be expected to be similar to that of reducing $F_{\rm MC}$ because the synaptic inputs into the BC and the GCs are reduced in both cases.

\begin{table*}
\caption{Parameters for the synaptic currents $I_R^{(GC,S)}(t)$ into the GC. The GCs receive the direct excitatory input from the entorhinal cortex (EC) cells, the inhibitory input from the HIPP cells, the excitatory input from the MCs, and the feedback inhibition from the BCs.
}
\label{tab:Synparm1}
\begin{tabular}{|c|c|c|c|c|c|c|}
\hline
Target Cells ($T$) & \multicolumn{6}{c|}{GC} \\
\hline
Source Cells ($S$) & \multicolumn{2}{c|}{EC cell} & HIPP cell & \multicolumn{2}{c|}{MC} & BC \\
\hline
Receptor ($R$) & AMPA & NMDA & GABA & AMPA & NMDA & GABA \\
\hline
$K_{R}^{(T,S)}$ & 0.89 & 0.15 & 0.12 & 0.05 & 0.01 & 25.0 \\
\hline
$\tau_{R,r}^{(T,S)}$ & 0.1 & 0.33 & 0.9 & 0.1 & 0.33 & 0.9 \\
\hline
$\tau_{R,d}^{(T,S)}$ & 2.5 & 50.0 & 6.8 & 2.5 & 50.0 & 6.8 \\
\hline
$\tau_{R,l}^{(T,S)}$ & 3.0 & 3.0 & 1.6 & 3.0 & 3.0 & 0.85 \\
\hline
$V_{R}^{(S)}$ & 0.0 & 0.0 & -86.0 & 0.0 & 0.0 & -86.0 \\
\hline
\end{tabular}
\end{table*}

\begin{table*}
\caption{Parameters for the synaptic currents $I_R^{(T,S)}(t)$ into the HIPP cell, MC, and BC. The HIPP cells receive the excitatory input from the EC cells, the MCs receive the excitatory input from the GCs, and the BCs receive the excitatory inputs from both the GCs and the MCs.
}
\label{tab:Synparm2}
\begin{tabular}{|c|c|c|c|c|c|c|c|c|}
\hline
Target Cells ($T$) & \multicolumn{2}{c|}{HIPP cell} & \multicolumn{2}{c|}{MC} & \multicolumn{4}{c|}{BC}\\
\hline
Source Cells ($S$) & \multicolumn{2}{c|}{EC cell} & \multicolumn{2}{c|}{GC} & \multicolumn{2}{c|}{GC} & \multicolumn{2}{c|}{MC} \\
\hline
Receptor ($R$) & AMPA & NMDA & AMPA & NMDA & AMPA & NMDA & AMPA & NMDA \\
\hline
$K_{R}^{(T,S)}$ & 12.0 & 3.04 & 1.4 & 0.25 & 0.38 & 0.02 & 0.74 & 0.04 \\
\hline
$\tau_{R,r}^{(T,S)}$ & 2.0 & 4.8 & 0.5 & 4.0 & 2.5 & 10.0 & 2.5 & 10.0 \\
\hline
$\tau_{R,d}^{(T,S)}$ & 11.0 & 110.0 & 6.2 & 100.0 & 3.5 & 130.0 & 3.5 & 130.0 \\
\hline
$\tau_{R,l}^{(T,S)}$ & 3.0 & 3.0 & 1.5 & 1.5 & 0.8 & 0.8 & 3.0 & 3.0 \\
\hline
$V_{R}^{(S)}$ & 0.0 & 0.0 & 0.0 & 0.0 & 0.0 & 0.0 & 0.0 & 0.0 \\
\hline
\end{tabular}
\end{table*}
In addition, in the present work, we considered the disynaptic inhibitory effect of the MCs on the GCs (i.e., disynaptic inhibition to the GCs, mediated by the BC).
On the other hand, we did not take into consideration the disynaptic effect of the HIPP cells on the GCs; we considered only their direct inhibition to the GCs.
The HIPP cells may disinhibit the BC \cite{BN2}, and then the inhibitory effect of the BC on the GCs becomes decreased, which leads to increase in the activity of the GCs. Thus, the disynaptic effect of the HIPP cells on the GCs, mediated by the BC, tends to increase the activity of the GCs, in contrast to the disynaptic inhibition of the MCs to the GCs. Hence, in future work, to study the disynaptic effect of the HIPP cells on the GC in the DG network would be useful.
Also, in the present work, we did not consider the direct EC input to the MCs \cite{Hilus2} and the input from the GCs to the HIPPs \cite{BN2}. For more complete network for the DG, it would be necessary to include these synaptic connections in future works.

\section*{Acknowledgments}
This research was supported by the Basic Science Research Program through the National Research Foundation of Korea (NRF) funded by the Ministry of Education (Grant No. 20162007688).

\appendix
\section{Parameter Values for The LIF Spiking Neuron Models and The Synaptic Currents}
\label{app:A}
In this appendix, we list three tables which show parameter values for the LIF spiking neuron models in Subsec.~\ref{subsec:LIF} and the synaptic currents in   Subsec.~\ref{subsec:SC}. These values are based on physiological properties \cite{Chavlis,Hilus3}. For the LIF spiking neuron models,  the parameter values for the capacitance $C_X,$ the leakage current $I_L^{(X)},$ and the AHP current $I_{AHP}^{(X)}$ are given in Table \ref{tab:Singleparm}.

The parameter values for the synaptic strength per synapse $K_{R}^{(T,S)}$ ($R:$ receptor, $S:$ pre-synaptic source population, and $T:$ post-synaptic target population), synaptic rising time constant $\tau_{R,r}^{(T,S)},$ synaptic decay time constant $\tau_{R,d}^{(T,S)},$ synaptic latency time constant $\tau_{R,l}^{(T,S)},$ and the synaptic reversal potential  $V_{R}^{(S)}$ for the synaptic currents into the GCs and for the synaptic currents into the HIPP cells, the MCs and the BCs are given in Tables \ref{tab:Synparm1} and \ref{tab:Synparm2}, respectively.

\end{document}